\documentclass[journal]{IEEEtran}

\title{Variable-Pitch-Propeller Mechanism Design, and Development of Heliquad for  Mid-flight Flipping and Fault-Tolerant-Control}
\author{Eeshan Kulkarni and Suresh Sundaram

\thanks{Eeshan (corresponding author) and Suresh are with the Department of Aerospace Engineering, Indian Institute of Science, Bangalore,
KA, 560012 India. e-mail: eeshank@iisc.ac.in}
}
\date{October 2023}

\usepackage{graphicx}
\usepackage{amsmath,eqnarray,amsthm}
\graphicspath{{Diagrams/}}
\usepackage{bm}
\usepackage{graphicx}
\usepackage{amssymb}
\usepackage{tikz}
\usepackage{subcaption}
\usepackage{array}
\usepackage{url}
\usepackage{textcomp}
\usepackage[font=small,labelfont=bf]{caption}
\usepackage{algorithmic}
\usepackage[ruled]{algorithm2e}

\usepackage{upgreek }
\usepackage{bm}

\usepackage{float}
\usepackage{tabularx}
\usepackage{adjustbox}
\graphicspath{{Diagrams/}}

\begin{document}

\maketitle

\begin{abstract}
    This paper presents the design of Variable-Pitch-Propeller mechanism and its application on a quadcopter called Heliquad to demonstrate its unique capabilities. The input-output relationship is estimated for a generic mechanism. Various singularities and actuator sizing requirements are also analyzed. The mechanism is manufactured, and the validated input-output relationship is implemented in the controller of Heliquad. Heliquad is controlled by a unified non-switching cascaded attitude-rate controller, followed by a unique Neural-Network-based reconfigurable control allocation to approximate nonlinear relationship between the control input and actuator command. The Heliquad prototype's mid-flight flip experiment validates the controller's tracking performance in upright as well as inverted conditions. The prototype is then flown in upright condition with only three of its working actuators. To the best of the authors' knowledge, the cambered airfoil propeller-equipped Heliquad prototype demonstrates full-attitude control, including yaw-rate, on three working actuators for the first time in the literature. Finally, the utility of this novel capability is demonstrated by safe recovery and precise landing post-mid-flight actuator failure crisis. Overall, the controller tracks the references well for all the experiments, and the output of the NN-based control allocation remains bounded throughout.
\end{abstract}

\section{Introduction} \label{intro_section}

 In the ever-expanding domain of Unmanned Aerial Vehicles (UAVs), the quadcopter design has emerged as a versatile and widely employed platform, showcasing mechanical simplicity, agility, and accessibility. Quadcopters find extensive applications across diverse fields, from aerial surveillance \cite{surv} and environmental monitoring \cite{enviro} to precision agriculture \cite{agri} and fire-fighting \cite{fire}, showcasing their adaptability in addressing a wide range of real-world challenges. Despite their versatility, traditional quadcopters with Fixed-Pitch-Propellers (FPP) face fundamental limitations in power efficiency, maneuverability, and actuator fault-tolerance, prompting the exploration of innovative solutions to overcome these constraints. The use of Variable-Pitch-Propellers (VPP) on all four actuators may solve some fundamental limitations of traditional FPP driven quadcopters. For example, it is shown in \cite{naoki2023mode} that the actuator redundancy in VPP systems may be exploited to optimize the power efficiency and thrust response.



\begin{figure}[t!]
 \begin{subfigure}{0.49\columnwidth}
  \includegraphics[width=\columnwidth]{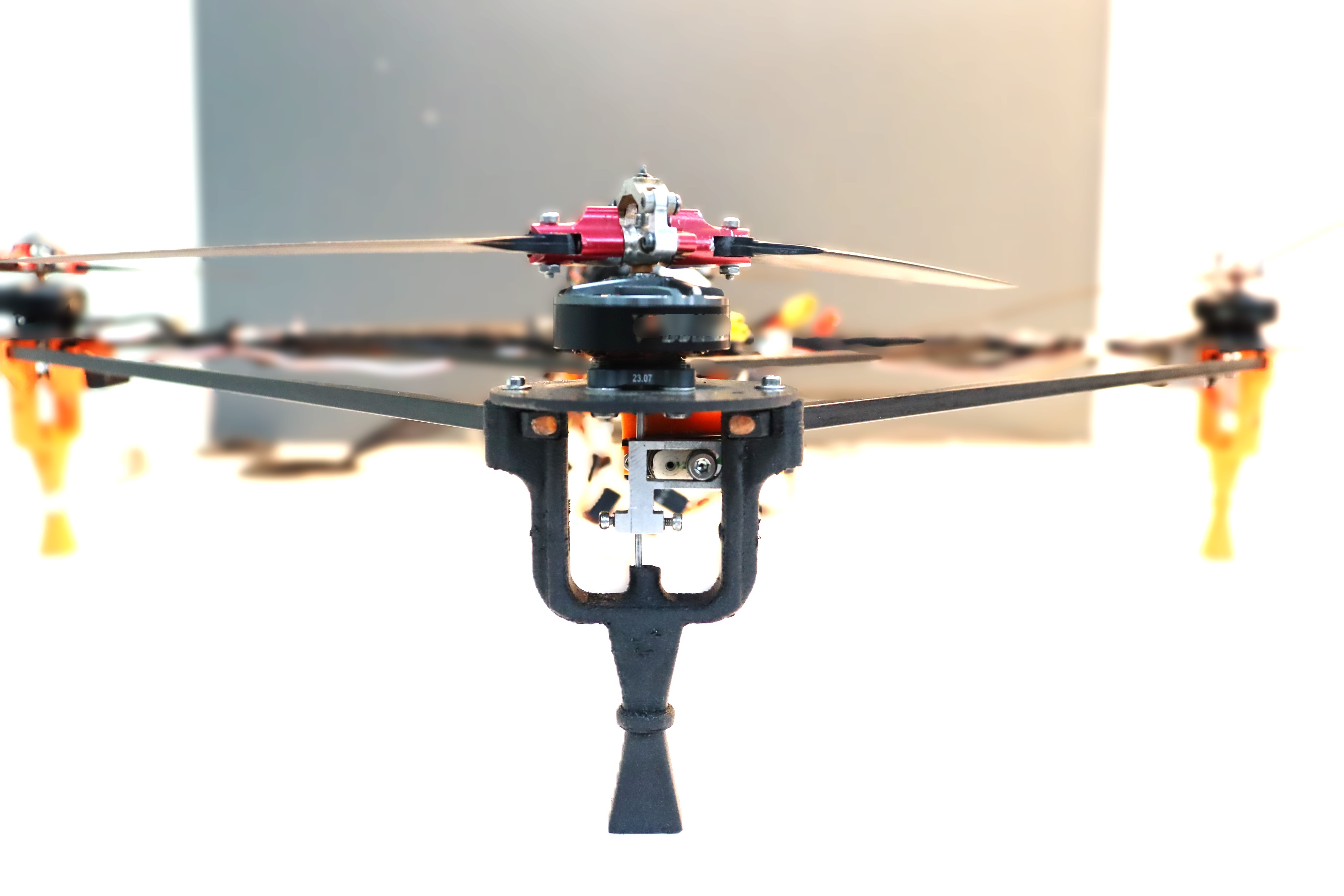} 
  \caption{ Variable-Pitch-Propeller actuator }
  \label{1st_page_a}
 \end{subfigure} 
 \hfill
 \begin{subfigure}{0.495\columnwidth}
  \includegraphics[width=\columnwidth]{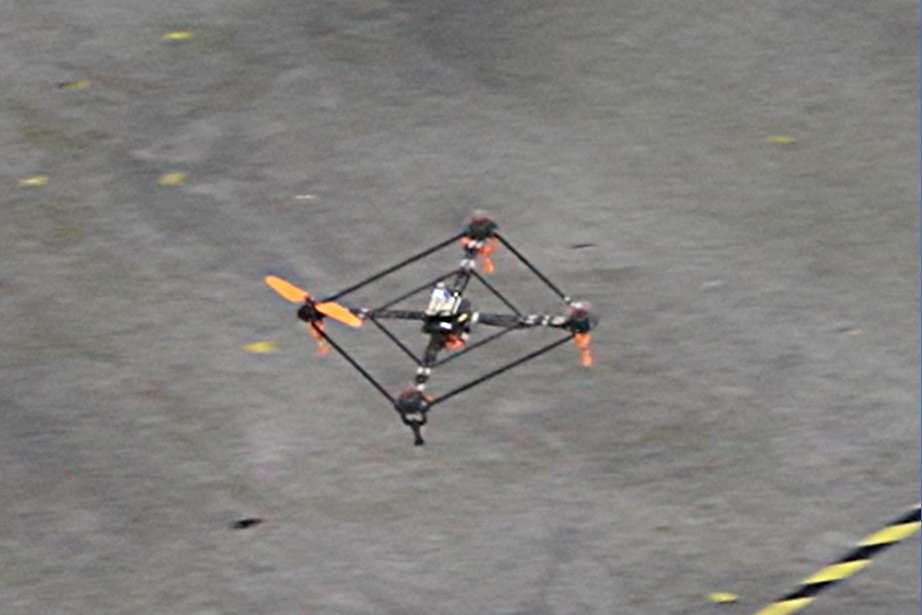}
  \caption{ Heliquad flying with a failed motor }
  \label{1st_page_b}
 \end{subfigure}
 
 \vspace{0.4cm}
 
 \begin{subfigure}{\columnwidth}
  \includegraphics[width=0.32\columnwidth]{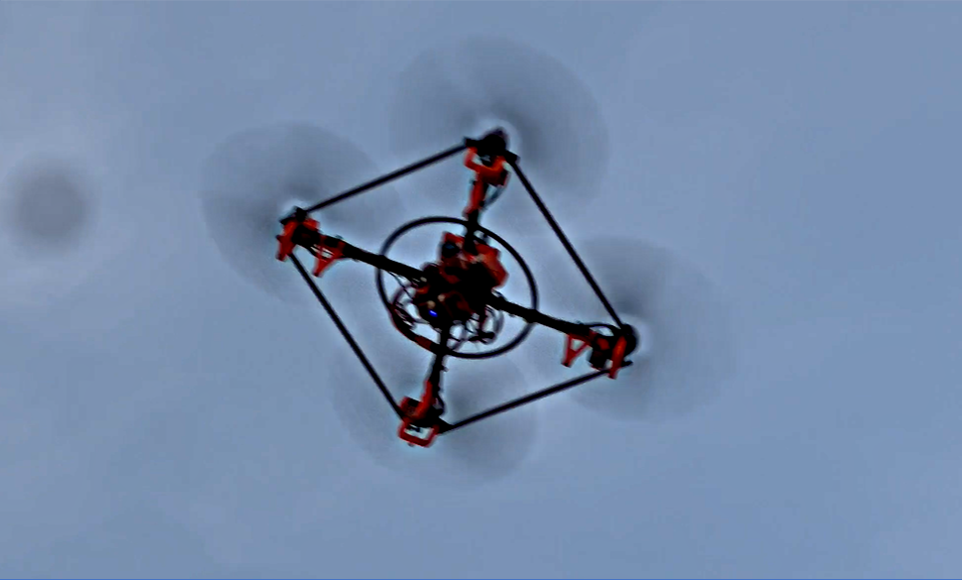}
  \hfill
  \includegraphics[width=0.32\columnwidth]{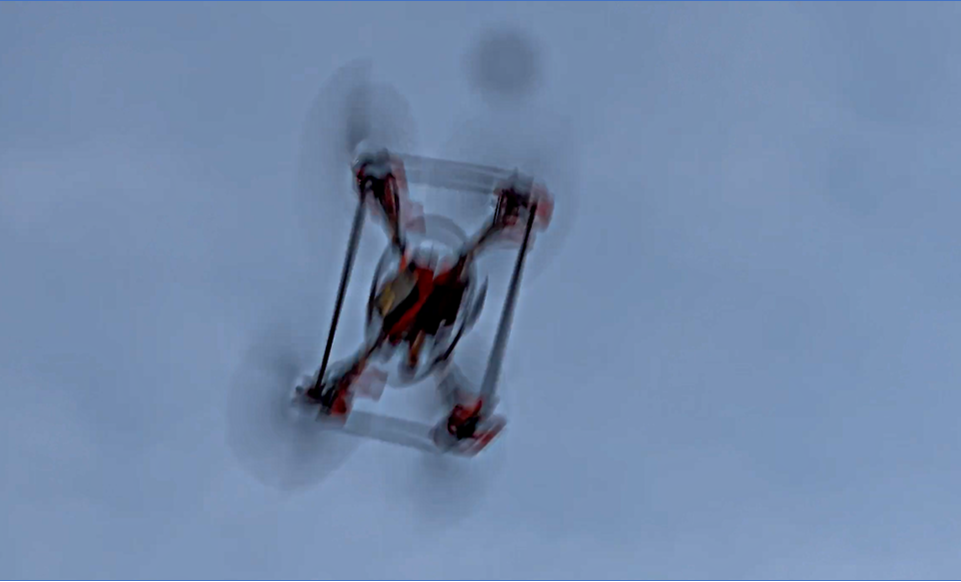}
  \hfill
  \includegraphics[width=0.32\columnwidth]{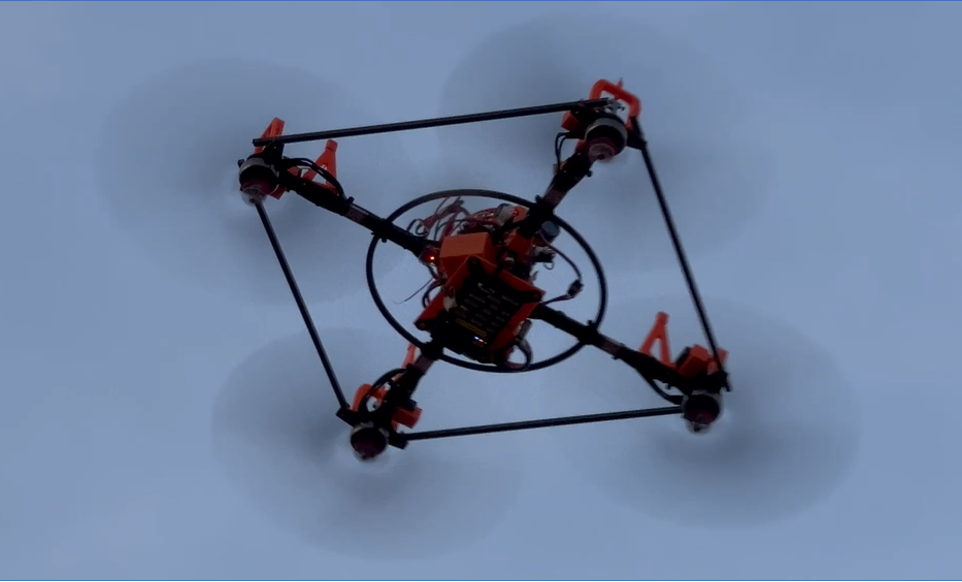}
  \label{1st_page_c}
  \caption{(L to R) Heliquad flipping from upright to inverted state mid-flight.}
 \end{subfigure}

 \caption{ Heliquad's Variable-Pitch-Propeller actuator can make it fly with full-attitude control despite the complete failure of an actuator. It can also perform mid-flight flips. The videos are included in the supplementary material.}
 \label{first_page_photo}
\end{figure}

Another interesting application to resolve the redundancy in VPP multirotors could be in the domain of actuator Fault-Tolerant-Control (FTC). Actuators on any multirotor may fail due to overheating, propeller strike, winding ingress, etc. The performance of VPP quadcopter under actuator loss of effectiveness wherein the actuator provides lesser control effort than expected is simulated in \cite{baldini_ftc}. Although the attitude references are tracked well, the system is not designed for a complete failure of an actuator. The pitch-stuck fault of a centrally powered VPP quadcopter where all the propellers are constrained to spin at the same Revolutions Per Minute (RPM) is investigated in \cite{ast_ftc}. It was shown in the simulations that after the fault, yaw and yaw-rates become uncontrollable. Such a reduced attitude control is also demonstrated under the complete failure of an actuator on FPP quadcopters \cite{jgcd_ftc},\cite{asme_ftc}. Full-attitude control under the complete failure of an actuator on a VPP quadcopter is predicted to be feasible using a cambered airfoil propeller \cite{kulkarni}.

The controller design in VPP-based UAV systems is complicated due to actuator redundancy and the highly nonlinear input-output relationship in propeller aerodynamics. The position and attitude controllers that are proven to work for FPP systems can also be used for controlling VPP systems \cite{lulea}, \cite{kawasaki2013muwa}. However, the control allocation methods may differ due to multiple actuator commands satisfying the desired thrust values. The control allocation method may depend on user requirements. For aggressive maneuvering, the motor RPM should be held constant at a higher value, and the propeller pitch angle alone should be controlled, whereas for optimizing input power, the pitch angle should be held constant at a higher value, and motor RPM can be controlled \cite{asme_ftc}. Five different control allocation methods are presented in \cite{krishna2022robust}. While these iterative methods are shown to work well in idealized numerical simulations, their performance, in reality, may not be as good due to inherent uncertainties in modeling and limited onboard processing power. The use of reinforcement learning-based controllers to undertake complex maneuvers autonomously on a VPP quadcopter is studied in \cite{wang2022aerobatic}. The controller is shown to execute the tic-toc maneuver in simulations with sufficient accuracy. However, the study is too restrictive as it considers the dynamics of a planar quadcopter. 

Another feature of the VPP multirotors is to fly upright as well as inverted. It is possible due to the actuator's ability to reverse the thrust vector. Utility-wise midflight flipping followed by sustained inverted flight may be used to expose the usually bottom-mounted payload towards the area above the multi-rotor. The midflight flipping and sustained inverted flight on a VPP quadcopter in an indoor environment is demonstrated in \cite{cutler_2015}. These maneuvers are beyond the dynamics of FPP quadcopters. It is worth noting that except \cite{cutler_2015}, no other novel VPP quadcopter in the literature has experimentally demonstrated any unique dynamic capability over traditional FPP quadcopters. However, actuator fault-tolerant control was not explored, and the control allocation nonlinearities were not addressed

The aforementioned VPP systems employ an additional mechanism to control the propeller pitch angle. Comprehensive insights into the pitch-controlling mechanism are necessary for any VPP system to determine the actuator input-output relationship, design scalability, singularity estimation, and actuator sizing. A fundamental investigation of the VPP mechanism design and its analysis has not yet been undertaken. This paper presents the design of a general Variable-Pitch-Propeller mechanism and its application on a quadcopter named Heliquad to demonstrate useful capabilities that are beyond the dynamics of traditional FPP quadcopters. Heliquad is a word coined by joining Helicopter and Quadcopter. The propeller pitch on the Heliquad can be controlled like a Helicopter's tail rotor (collective pitch). However, it can be maneuvered like a quadcopter (by differential thrust). The mechanism to control the propeller pitch is similar to a variant of four-bar chain. The input-output relationship of this mechanism is estimated and later validated for its use in the controller. As the mechanism is closed-loop, its singular configurations are estimated, and the link parameters are chosen to avoid singularities in the operating pitch angle range. To adequately size the actuator, external torque acting on it due to the propeller's pitching moment is also estimated. The Heliquad is controlled by a unified non-switching cascaded attitude-rate controller followed by the reconfigurable control allocation. Neural networks are employed in the control allocation to handle nonlinearities in propeller aerodynamics. A prototype is built to validate the utility of the mechanism and performance of the controller. The mid-flight flipping experiments are conducted to validate the controller's tracking performance in upright and inverted conditions. Then, by switching off one motor, the prototype is flown with full-attitude control, including yaw-rate on only three working actuators. Finally, by switching off one of the motor midflight, the safe recovery and precise landing of the Heliquad prototype is demonstrated.  To the best of the authors's knowledge, this paper demonstrates full-attitude control, including the yaw-rate tracking, on three working actuators for the first time in VPP quadcopter literature. In this case, the role of a cambered airfoil in the propeller blades is pivotal to ensure the feasibility of the hover equilibrium point.  The readers are referred to \cite{kulkarni} for detailed aerodynamic analysis on this topic. Note that the hover equilibrium point does not exist for the Heliquad with two or three completely failed actuators. However, reduced attitude techniques may be used to land safely after such an event.

The paper is organized as follows: The major notations used in the paper are given in section \ref{notation_section}. The VPP mechanism design and analysis is provided in section \ref{vpp_mechanism_section}. The Heliquad's dynamics and unified attitude controller are presented in section \ref{heli_dc_section}. The Heliquad prototype's description and the experimental results are shown in section \ref{heli_results_section}. The conclusion and future works are given in section \ref{conclusion_section}.

\section{Notation} \label{notation_section}

Some general notations used in the paper are presented in Table.\ref{notation_table}. Lowercase nonbold symbols represent the scalars.  Vectors are described by bold lowercase symbols, and matrices are given in uppercase bold symbols.

\begin{table}[h!]
    \centering
    \begin{tabular}{|m{2.4cm}|m{5.3cm}|}
    \hline
    \textbf{Symbol} & \textbf{Meaning} \\
    \hline
    $\{i\}$ & Co-ordinate system $i$ with the origin at $O_i$\\
    \hline
      $\eta_i , \ \forall  i \in\{1,2,3\} $ & Mechanism joint angles \\
      \hline
      $\gamma$, $\gamma_i$ & Propeller pitch angle, propeller pitch angle for $i^{th}$ actuator on Heliquad.\\
      \hline
       $l_i , \ \forall  i \in \{1,2,3,4\} $ & Length of link $i$ \\
       \hline
       ${}^{k}\bm{x}_{ij} \in \mathbb{R}^{3}$ & Displacement vector between $\{i\}$ and $\{j\}$ expressed in $\{k\}$\\
       \hline
       ${}^i\bm{R}_{j}\in \mathbb{R}^{3\times3}$  & Rotation matrix that rotates $\{i\}$ to $\{j\}$\\
       \hline
       $T_i,\tau_i \ \forall i \in {1,2,3,4} $  & Thrust/Torque generated by $i^{th}$ propeller \\
       \hline
       $T_\Sigma$ & Collective thrust, $\sum_{i=1}^{4} T_i$ \\
       \hline
       $\bm{I} \in \mathbb{R}^{3\times3}$ & Moment of inertia matrix for Heliquad \\
       \hline
       $\bm{{}^B\omega}_{ij}$ & Body-fixed angular velocity of $\{j\}$ with respect to $\{i\}$ \\
       \hline
       $\bm{m_B} \in \mathbb{R}^{3}$ & Moments expressed in the body frame \\
       \hline
       $x_d$  & Desired value of $x$. \\
       \hline
       $\dot{x}$ & Time derivative of the individual element of $x$. \\
       \hline

   \end{tabular}
   \caption{Notations}
    \label{notation_table}
\end{table}

\section{Variable-Pitch-Propeller Mechanism Design and Analysis} \label{vpp_mechanism_section}

In this section, the Variable-Pitch-Propeller mechanism design and analysis is presented. The relationship between input-output angles is derived. Then, singularities in the mechanism are estimated. This section ends with the derivation of an expression for the minimum required torque for the propeller pitch-controlling actuator. In VPP systems, an extra actuator is used per rotor to control the propeller pitch angle precisely. This actuator, typically a servo-motor, is in addition to the other actuator(s) that rotates the propeller blades. On multi-rotor systems, either a centralized source such as an Internal-Combustion (IC) engine is used to spin all the propellers at the same RPM, or every propeller is rotated by a separate source such as an electric motor, typically a Brushless Direct Current (BLDC) motor. 

\begin{figure}[t!]
 \begin{subfigure}{0.45\columnwidth}
  \includegraphics[width=\columnwidth]{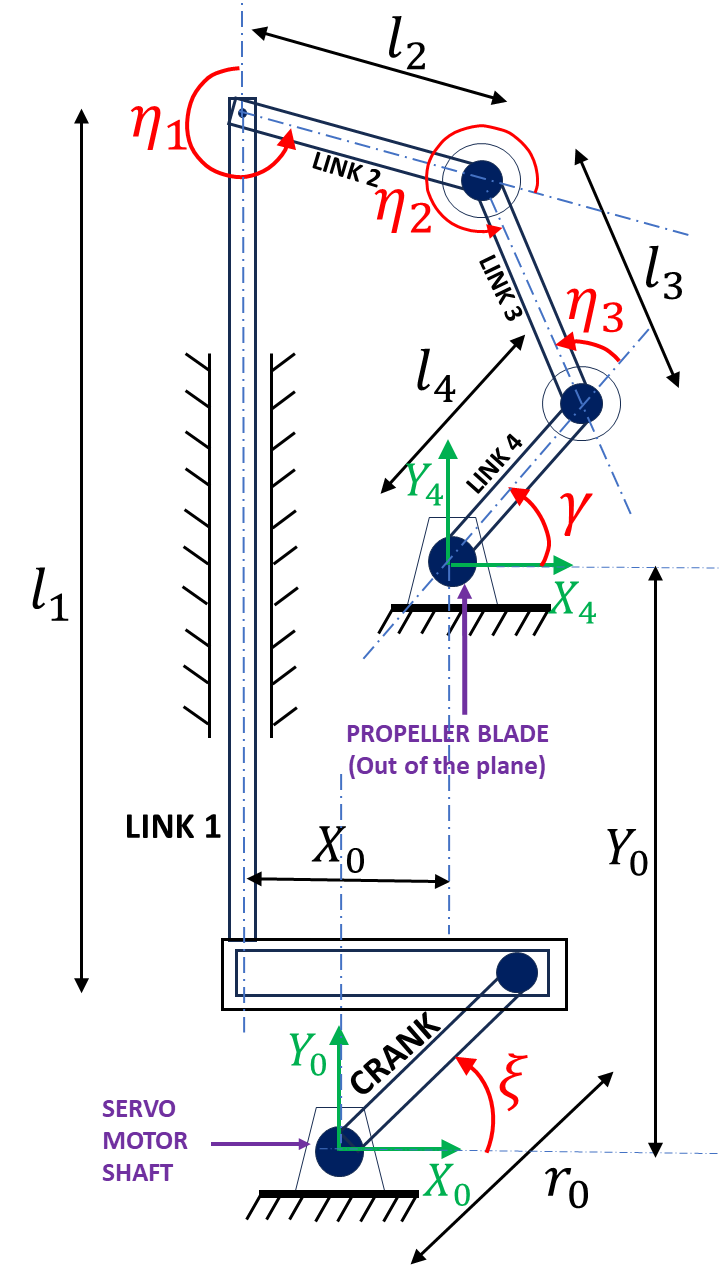} 
  \caption{Mechanism schematic. }
  \label{mechanism_schematic_a}
 \end{subfigure} 
 \hfill
 \begin{subfigure}{0.45\columnwidth}
  \includegraphics[width=\columnwidth]{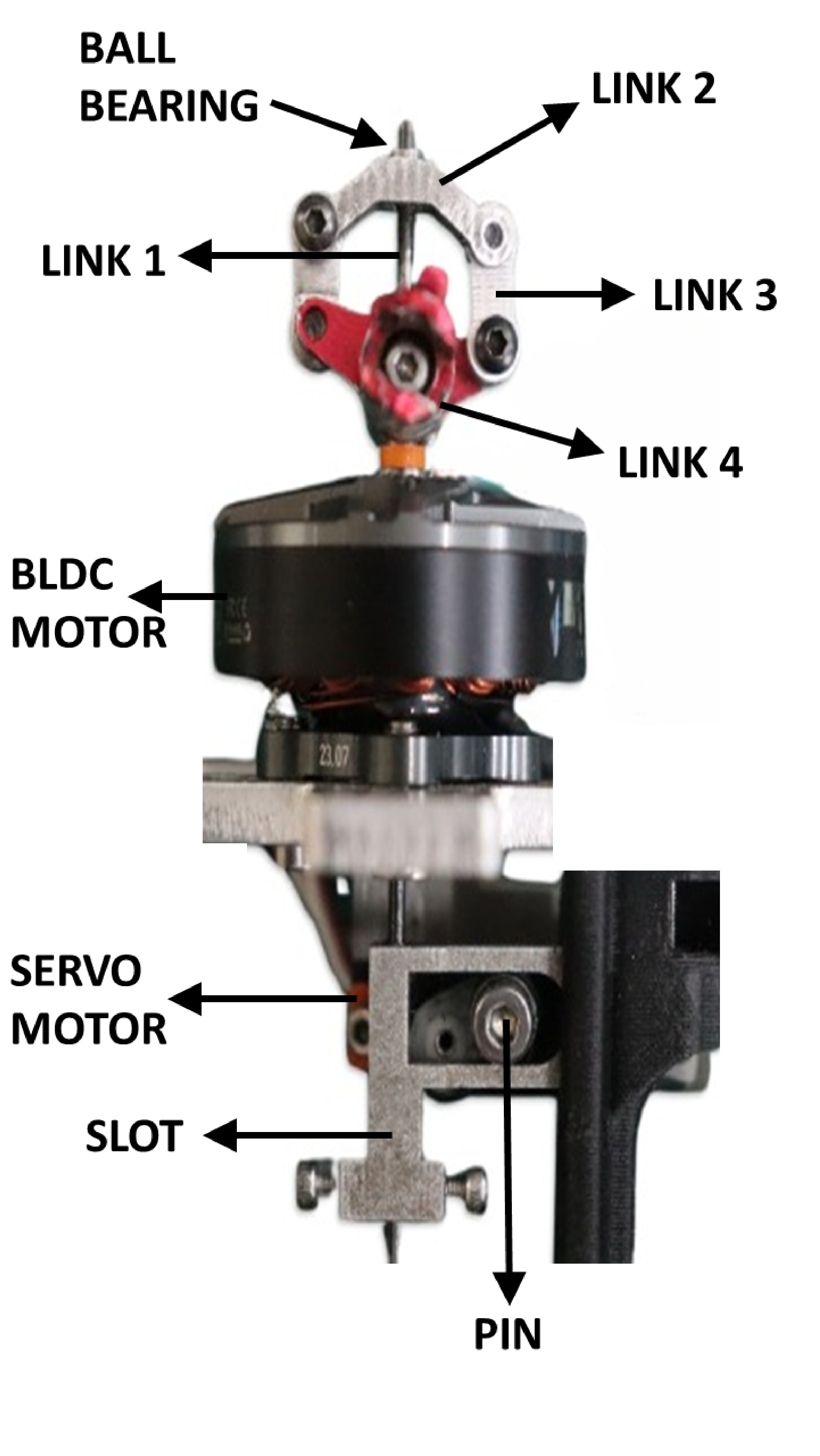}
  \caption{ Heliquad's actuator }
  \label{mechanism_schematic_b}
 \end{subfigure}
 \caption{ (a) The mechanism schematic used for analysis. $r_0$ is the crank length. (b)  Actual VPP actuator mounted on Heliquad. The servo-motor is mounted behind the slot. The actuator has two blades. Another link similar to Link 4 is behind it. }
 \label{mechanism_schematic}
\end{figure}

The VPP mechanism should ideally have one Degree-Of-Freedom (DOF) and be mechanically as simple as possible. One such generic mechanism is shown in Fig.\ref{mechanism_schematic}. The same is implemented in Heliquad.  The servo-motor precisely rotates the input link, which is connected to a series of subsequent links ending in the output link to which the propeller blade is fixed. This results in a planar closed-loop (parallel) mechanism. For clarity, in Fig.\ref{mechanism_schematic_a}, only a single propeller blade is shown. Multiple propeller blades can be mounted on an individual mechanism based on the requirement. The mechanism controls the collective pitch of the propeller. It means all the blades will have the same pitch angle at any given time, irrespective of azimuth. The Heliquad's mechanism has two propeller blades.

In Fig.\ref{mechanism_schematic_a}, Link 1 is constrained to move vertically. In practice, Link 1 will pass through the hollow shaft of the rotating BLDC motor as shown in Fig.\ref{mechanism_schematic_b}. A ball bearing is fixed to the top end of Link 1. Link 2 is fixed on the rotating outer race of the bearing. For simplicity in analysis, the mechanism is assumed to be planar by considering Link 2 to be fixed in the plane ($\eta_1$ is constant). Link 2 is connected to Link 3 by a revolute joint. Link 3 is further connected to Link 4 by another revolute joint. Link 4 is grounded on the other end. In practice, Link 4 is fixed on the outside of the hollow shaft of the rotating BLDC motor.  Therefore, Link 1 will remain stationary with respect to the ground, and Links 2, 3, and 4 will rotate at the motor's RPM. The propeller blade is rigidly connected to Link 4. Depending on the number of propeller blades, Link 2 may be connected to multiple chains of links similar to Links 3 and 4. To avoid imbalance, these links must have exact dimensions and should be arranged symmetrically in the top plane. Hence, for $n_B$ propeller blades, there will be $n_B$ links similar to Link 3 and $n_B$ links similar to Link 4. 

To change the propeller pitch angle, the servo-motor needs to control the verticle motion of Link 1. This paper uses a pin-slot mechanism to convert servo-motor rotation to the linear motion of Link 1. Other mechanisms, similar to serial 2R manipulators can also be explored. As shown in Fig. \ref{mechanism_schematic}, a crank is connected to the servo-motor shaft, and a pin is attached to the other end of the crank. The pin has a sliding fit in the slot. The slot is further rigidly connected to Link 1. The overall mechanism is similar to the RPRR variant of the 4-bar chain, with one active joint variable $\xi$ (servo-motor angle with respect to horizontal) and three passive joint variables $\eta_2$, $\eta_3$, and $\gamma$,  the last one being the propeller pitch angle.

\subsection{VPP mechanism input-output relationship}

The mathematical relation between $\gamma$ (output) and $\xi$ (input) is essential for mechanism evaluation and controller design. In this section, the direct kinematics problem of the mechanism is solved. As the mechanism is closed-loop in nature, to derive the constraint equations, the joint between Link 2 and Link 3 is assumed to be broken. Then, the distances from both sides around the loop to the broken joint from $\{ 4 \}$ are equated, resulting in,

\begin{equation}
l_4 \ cos(\gamma) \ + \ l_3 \ cos(\gamma + \eta_3) = -X_0 \ + \ l_2 \ cos(\frac{\pi}{2} + \eta_1)    
\label{loop_closureX}
\end{equation}
\begin{multline}
l_4 \ sin(\gamma) \ + \ l_3 \ sin(\gamma + \eta_3) = -Y_0 \ + \ r_0\ sin(\xi) \ + \ l_1  + \\ \ l_2 \ sin(\frac{\pi}{2} + \eta_1)    
\label{loop_closureY}
\end{multline}
Eq.\ref{loop_closureX} and Eq.\ref{loop_closureY} are the constraint equations of the mechanism. Note, the equations do not include passive variable $\eta_2$. Right-Hand-Side (RHS) of Eq.\ref{loop_closureX} is constant. RHS of Eq.\ref{loop_closureY} consists of a constant and a term with an active variable (actuator input). Squaring and adding both equations gives
\begin{equation}
    cos\gamma\ cos\delta + sin \gamma \ sin\delta =\frac{A^2 \ +\ B^2 \ - \ l_4^2\ -\ l_3^2}{2l_4l_3}
    \label{ccss}
\end{equation}
where,
\begin{equation}
\begin{split}   
 A\ &= -X_0 \ + \ l_2 \ cos(\frac{\pi}{2} + \eta_1) \\
 \ B \ &= \  -Y_0 \ + \ r_0\ sin\xi \ + \ l_1  + \ l_2 \ sin(\frac{\pi}{2} + \eta_1) \\
 \delta &= \ \gamma \ + \ \eta_3
 \end{split}
\end{equation}
The expression for passive variable $\eta_3$ can be derived from Eq.\ref{ccss}
\begin{equation}\label{eta3}
    \eta_3 \ = \ cos^{-1} (\ C\ )
\end{equation}
where,
\begin{equation}
    C \ = \ \frac{A^2 \ + \ B^2 \ - \ l_4^2\ -\ l_3^2}{2l_4l_3}
\end{equation}
Substituting Eq.\ref{eta3} in Eq.\ref{loop_closureX}, and further simplification results in
\begin{equation}\label{PQeta}
 P \ cos\ \gamma \ + \ Q \ sin\ \gamma \ = \ A   
\end{equation}
where
\begin{equation}
\begin{split}
    P \ &= \ l_4 \ + \ l_3 \ C \\
    Q \ &= \ -l_3 \ \sqrt{1\ - \ C^2} 
\end{split}
\end{equation}
Solving for $\gamma$ in Eq.\ref{PQeta} gives the input-output relation
\begin{equation}\label{iorel}
\gamma \ = \ tan^{-1} \left(\frac{Q}{P}\right) \ + \ cos^{-1} \left(\frac{A}{\sqrt{P^2+Q^2}}\right)
\end{equation}
In the above equation, $P$ and $Q$ are functions of $\xi$. A detailed derivation of Eq.\ref{iorel} is provided in the supplementary material. Eq.\ref{iorel} can be iteratively solved for different $\xi$'s, and a curve may be fitted to approximate the relation for implementation in the controller.

\subsection{Kinematic singularities in the mechanism}

Like any other closed-loop kinematic chain, different kinds of singularities are encountered in the VPP mechanism. These singularities can be classified into three groups \cite{singularities}. In the first kind, called the dwell, the concerned link is at the dead point. In the VPP mechanism, it corresponds to the propeller pitch angle not changing despite finite servo-motor rotation.  In the second kind of singularity, the mechanisms gain one or more DOFs. Here, the passive joints may move with the actuators locked. Therefore, the pitch angle may change while the servo-motor shaft is fixed. The third kind combines the first two singularities and can occur only for special link parameters. In the VPP mechanism, these singularities are not desired. The focus of this section will be to identify and avoid them. Hence, the third singularity is not discussed.

Differentiating Eq.\ref{loop_closureX} and Eq.\ref{loop_closureY} with respect to time and subsequently separating active and passive joint variables yields,

\begin{equation}
  \bm{K}
    \begin{bmatrix}
        \Dot{\gamma}\\
        \Dot{\eta_3}
    \end{bmatrix}
   \  +
 \  \bm{k}
    \Dot{\xi} \ = \ \bm{0_{2\times1}}
\end{equation}

where,
\begin{multline} \label{Kk+}
 \bm{K} \ = 
   \begin{bmatrix}
        -l_4 sin\gamma -  l_3  sin(\gamma + \eta_3) & -l_3 sin(\gamma \ + \ \eta_3) \\
        l_4 cos(\gamma) + l_3 cos(\gamma + \eta_3) & l_3 cos(\gamma + \eta_3)
    \end{bmatrix}, \\
    \bm{k} \ = 
      \begin{bmatrix}
        0\\
        -r_0 cos(\xi)
    \end{bmatrix}
\end{multline}
The rate of change of passive variables as a function of actuator variable can be written as

\begin{equation}
    \begin{bmatrix}
        \Dot{\gamma}\\
        \Dot{\eta_3}
    \end{bmatrix}
   \ = \ - \ \bm{K}^{-1} \ \bm{k} \ \Dot{\xi}
\end{equation}
\\
\begin{multline}\label{kinvk}
\bm{K}^{-1} \ \bm{k}  \ = \ 
\big[
 \frac{r_0\,\sin\left(\gamma +\eta _{3}\right)\,\cos\xi }{l_{4}\,\sin\eta _{3}}  \quad \\
 -\frac{r_0\,\cos\xi \,\left(l_{3}\,\sin\left(\gamma +\eta _{3}\right)+l_{4}\,\sin\gamma\right)}{l_{3}\,l_{4}\,\sin\eta _{3}} 
\big] ^{T} \
\end{multline}
For the first kind of singularity to occur, the first row of $\bm{K}^{-1} \ \bm{k}$ should be null. As seen from Eq.\ref{kinvk}, Link 4 will instantaneously dwell when $\xi$ is $\pm \frac{\pi}{2}$, which is intuitive as the Link 1 will not move vertically. In practice, if for the desired $\gamma$, $\xi$ is near $\frac{\pi}{2}$, $r_0$ may be increased.

%

The second kind of singularity occurs when $det(\bm{K}) = 0$. From Eq.\ref{Kk+}, after simplification, one can write
\begin{equation}\label{detK}
\begin{split}
    det(\bm{K}) \ &= \ l_3 l_4 \ [ \ cos\gamma \ \sin(\gamma + \eta_3) - sin\gamma \ cos(\gamma + \eta_3) \ ] \\ 
    \\
    &= \ l_3 l_4 \ sin \eta_3
\end{split}
\end{equation}
Eq.\ref{detK} will be zero if $\eta_3$ is 0 or $n \pi$, where $n$ is an integer.  Due to the definition of $\eta_3$ (see Fig.\ref{mechanism_schematic}), $n$ must be 1. Therefore, this singularity will occur when passive links 3 and 4 are in a straight line \footnote{The demonstration of the singularity is shown in the supplementary video}. After substituting the singular value of $\eta_3$ in Eq.\ref{loop_closureX}, the singular value of pitch angle ($\eta_s$) can be written as
\begin{equation}
    \eta_{s} \ = \ \frac{A}{l_4 \pm l_3}
\end{equation}
At $\gamma = \eta_s$, the joint connecting Link 3 and 4 will rotate, changing the propeller pitch angle even when the servo-motor is locked. This is undesirable as forces generated by propellers will change suddenly, risking the stability of the entire system. As this singularity may occur within the workspace of the mechanism, the link lengths should be designed in a way to prevent $\eta_s$ from being in the operational propeller pitch angle ranges. In practice, ideally, the value of $\eta_s$ could be greater than the value of $\gamma$ where propeller airfoil stalls.

\subsection{Actuator sizing}
The servo-motor sizing for the mechanism is vital for precise control of Heliquad. As the mechanism rotates, the propeller will exert moments on Link 4. In these conditions, the servo-motor should be powerful enough to enable accurate adjustments in propeller pitch angle. Hence, to size the servo-motor adequately, an expression for the minimum required torque to ensure static equilibrium is required. Similar to the statics of any generic manipulator, the actuator torque required to maintain static equilibrium is calculated as

\begin{equation}
    \bm{\tau_{act}} \ = \ \bm{J}^{T} \  \bm{f_{ext}}
\end{equation}
where $\bm{\tau_{act}}$ is the actuator torque vector, $\bm{J}$ is the jacobian matrix and $\bm{f_{ext}}$ is the generalized moment vector. Substituting the value of $\bm{J}$ from Eq.\ref{kinvk}, the minimum servo-motor torque required to maintain equilibrium can be written as
\begin{equation}\label{prop_moment}
    \tau_s \ =  \ \frac{M_{prop} \ r_0 \ l_3 \ sin(\gamma \ + \eta_3) \ cos\xi}{l_4 \ sin\eta_3}
\end{equation}
Where, $M_{prop}$ is the moment generated by the rotating propeller. $M_{prop}$ can be estimated by  Blade-Element-Momentum-Theory (BEMT) \footnote{BEMT equations are provided in the supplementary material} \cite{leishman}.

\section{Heliquad dynamics and unified control} \label{heli_dc_section}

In this section, the dynamics of the Heliquad is formulated, and a unified controller is designed for mid-flight flipping and flight with three working actuators.
\subsection{Heliquad dynamics}
\begin{figure}
    \centering
    \includegraphics[width=0.75\columnwidth]{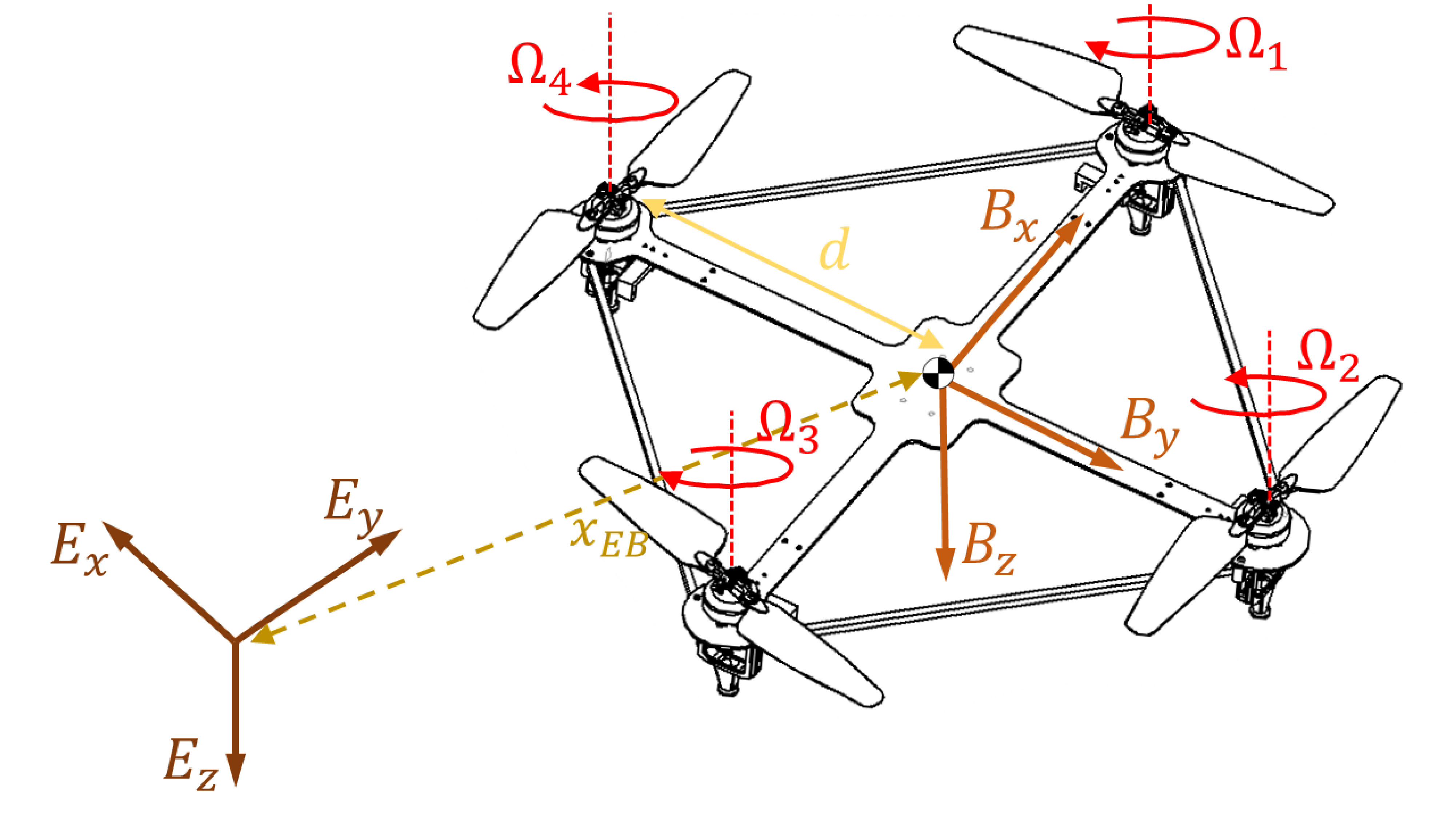}
    \caption{Frames of reference. $\{E\}$ is the inertial and $\{B\}$ is the body frame of refernce. Subscript of $\Omega$ represents actuator number.}
    \label{heli_layout}
\end{figure}

To gain deeper insights into the capabilities of Heliquad, it is essential to understand its dynamics. The layout of Heliquad is provided in Fig.\ref{heli_layout}. 
It utilizes a '$+$' frame with its actuators fixed along the planar body axes. The equations of motion for the Heliquad can be written as
\begin{equation} \label{eom}
    \begin{split}
       & m \ {}^{\bm{E}}\bm{\Ddot{x}_{EB}} \ = \ {}^E\bm{R}_{B} \ [ 0 \quad 0 \quad T_\Sigma]^T \\
       \\
        & \bm{I} \ {}^B\Ddot{\bm{\omega}}_{EB} \ + \ {}^B\bm{\omega}_{EB} \times \bm{I} \ {}^B \bm{\omega}_{EB} \ = \ \bm{m_B}
    \end{split}
\end{equation}

where, 
\begin{equation} \label{veemap}
    \bm{\omega}_{EB} \ =  \ s( {}^E\bm{R}_{B}^T \ {}^E\dot{\bm{R}}_{B} )
\end{equation}
$m$ is the mass of Heliquad. In Eq.\ref{veemap}, $s(.) : \mathbb{R}^{3\times3} \rightarrow \mathbb{R}^{3 \times 1}$ is the $vee$ map that maps the skew-symmetric matrix to the angular velocity vector. The dynamics of the Heliquad is similar to the standard fixed-pitch quadcopters. However, for Heliquad, $T_\Sigma$ in Eq.\ref{eom} can take negative values as well. This is possible due to its ability to generate thrust in both directions with respect to the body frame of reference. The control input for the Heliquad  can be written as
\begin{equation} \label{mixer_matrix}
    \begin{bmatrix}
        m_{Bx}\\
        m_{By}\\
        m_{Bz}\\
        T_\Sigma \\
    \end{bmatrix}
     =  \begin{bmatrix}
        0 & -d & 0 & d\\
        d & 0 & -d & 0\\
        -k_{\tau} (\gamma_1) & k_{\tau} (\gamma_2) & -k_{\tau} (\gamma_3) & k_{\tau} (\gamma_4)\\
        1 & 1 & 1 & 1\\
         \end{bmatrix}
         \begin{bmatrix}
             T_1 \\
             T_2\\
             T_3\\
             T_4\\
         \end{bmatrix}
\end{equation}
In the above equation, $k_\tau(\gamma_i)$ is a positive quantity defined as
\begin{equation}
    k_\tau(\gamma_i) = \big| \ \frac{\tau_i}{T_i} \ \big|
\end{equation}
where $\tau_i$ and $T_i$ are also the functions of $\gamma$. The matrix in Eq.\ref{mixer_matrix} is the control allocation matrix, and the same will be used in the controller design of the Heliquad for upright as well as inverted flight. 

\subsection{Controller design}
\begin{figure}[htbp]
    \centering
    \includegraphics[width=\columnwidth]{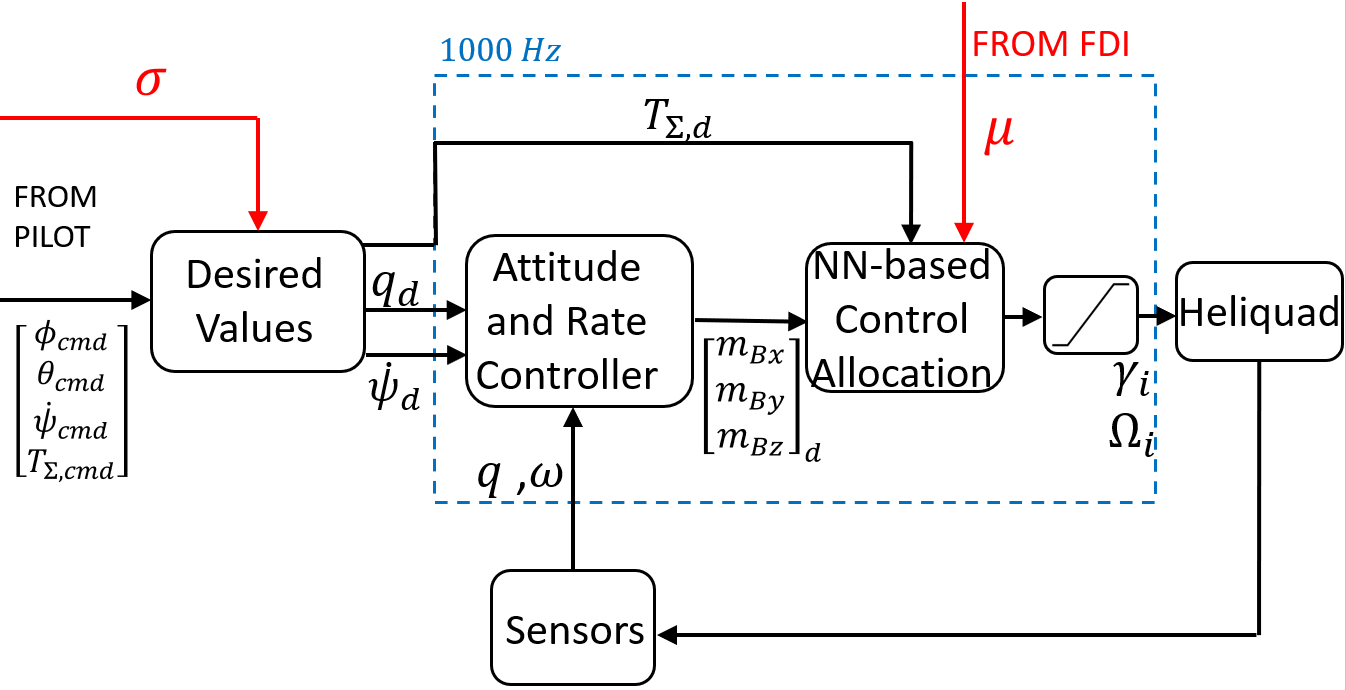}
    \caption{Control loop schematic. Fault Detection and Isolation (FDI) method is not discussed in this paper.}
    \label{control_schematic}
\end{figure}

In addition to regular upright flight, the Heliquad exhibits various dynamic capabilities such as mid-flight flipping, sustained inverted flight, and flight on three actuators. The controller should be able to stabilize and track the attitude throughout its flight envelope. Towards it, a unified non-switching attitude controller is developed. The following points describe the working of the control loop.
\begin{itemize}
    \item The upright or inverted state of the Heliquad is determined by a boolean variable $\sigma(t)$. $\sigma(t) = 0$  for upright and $\sigma(t)=1$  for inverted flight. Switching the values mid-flight will cause the Heliquad to flip 180 degrees and maintain the commanded attitude.
    \item The bounded roll ($\phi_{cmd}$), pitch ($\theta_{cmd}$), and yaw-rate ($\Dot{\psi}_{cmd}$) values are commanded as the reference for the controller to track. Note, the commanded values are represented in the frame with its origin fixed at $O_B$ and rotated by $\psi$ with respect to $\{E\}$. Therefore, the qualitative behavior of Heliquad shall be the same irrespective of whether it is in an upright or inverted flight. For example, a positive roll will make it move toward its right in both cases.
    \item Fault Detection and Isolation (FDI) routine outputs a parameter $\mu \in \{0 ,1,2,3,4\}$ that represents the index of the completely failed actuator. $\mu =0$ means all the actuators are working properly. The FDI method is not discussed in this paper. However, existing methods similar to \cite{mueller_fdi} may be used to identify the fault.
\end{itemize}
The control layout is shown in Fig.\ref{control_schematic}.

To disambiguate upright and inverted flight, the commanded values from the pilot are converted to the desired values depending on $\sigma (t)$ 
\begin{equation}\label{cmd_d_eq}
    \begin{split}
        \begin{bmatrix}
            \phi_d \\
            \theta_d\\
            \dot{\psi}_d\\
        \end{bmatrix}
        \ &= \ \sigma (t) \begin{bmatrix}
            \pi\\
            0\\
            0\\
        \end{bmatrix}
        \ + \ \begin{bmatrix}
            \phi_{cmd}\\
            \theta_{cmd}\\
            \dot{\psi}_{cmd}\\
        \end{bmatrix} \\
       T_{\Sigma,d} \ &= \ (-1)^{ 2-\sigma(t)} \ T_{\Sigma,cmd}
    \end{split}
\end{equation}
In Eq.\ref{cmd_d_eq}, it is considered that the Heliquad will flip along the roll axis. The flip can be performed about pitch axes by adding $\pi$ to $\theta_{cmd}$ instead. As the Heliquad is expected to perform aggressive large-angle maneuvers during mid-flight flipping, singularity-free representation of the attitude is essential. Hence, the quaternions are used in control calculations. The desired Euler angles in Eq.\ref{cmd_d_eq} can be converted to the desired quaternion as 
\begin{equation}\label{eul2q}
   \bm{q_d} =\left[\begin{array}{l}
c (\phi_d / 2) c (\theta_d / 2) c (\psi / 2)+s (\phi_d / 2) s (\theta_d / 2) s (\psi / 2) \\
s (\phi_d / 2) c (\theta_d / 2) c (\psi / 2)-c (\phi_d / 2) s (\theta_d / 2) s (\psi / 2) \\
c (\phi_d / 2) s (\theta_d / 2) c (\psi / 2)+s (\phi_d / 2) c (\theta_d / 2) s (\psi / 2) \\
c (\phi_d / 2) c (\theta_d / 2) s (\psi / 2)-s (\phi_d / 2) s (\theta_d / 2) c (\psi / 2)
\end{array}\right]
\end{equation}
Note, the yaw-rate is directly commanded to the rate controller. Therefore, in Eq.\ref{eul2q}, it is always assumed that $\psi_d = \psi$. 

\subsubsection{Attitude and rate controllers}

The rate controller is cascaded with the attitude controller. The attitude controller takes the desired and actual quaternion as the input and outputs the desired angular velocity. The rate controller takes in the desired and actual angular velocity and outputs the desired moments that have to be generated in the body frame. This cascaded control architecture is similar to open-source PX4 autopilot \footnote{\url{https://docs.px4.io/main/en/flight_stack/controller_diagrams.html}}. The attitude controller is given by
\begin{equation}
    \bm{\omega_{d}}' \ = \ sgn(q_{e,0}) \ \bm{K_a}  \ \bm{q_{e,1:3}}
\end{equation}
Where $\bm{K_a}$ is the diagonal matrix of controller gains and $\bm{q_e} = \bm{q^{-1}}\bm{q_d}$. The controller is shown to be globally asymptotically stable in \cite{eth_control}. It is also shown that the system reaches $\bm{q_d}$ with minimal rotation. Therefore, the Heliquad may not always flip in the same direction. The desired yaw-rate setpoint is added to $\bm{\omega_d}'$ and then passed to the rate controller.
\begin{equation}
    \bm{\omega_d} \ = \ \bm{\omega_d}' \ + \ [0 \quad 0 \quad \ \dot{\psi_d}]^T
\end{equation}
The angular velocity is controlled by a standard Proportional-Integral-Derivative (PID) controller.
\begin{equation}
    \bm{m_B} \ = \ \bm{K_p} \bm{\omega_e} \ + \ \bm{K_I} \int\bm{\omega_e} + \ \bm{K_D} \dot{\bm{\omega_e}} 
\end{equation}
where, $\bm{K_p}$, $\bm{K_I}$, and $\bm{K_D}$ are the tunable gains.

\subsubsection{Neural Network Based Control Allocation}
The control allocation maps $\bm{m_B}$ and $T_{\Sigma,d}$ to the actuator commands $\Omega_i$'s and $\gamma_i$'s. The desired collective thrust and body moments are first converted to the desired thrusts for individual actuators. Then a Neural-Network (NN) is used to map thrust to the actuator commands. The first step of a generalized control allocation is given by 
\begin{equation}\label{CA}
\bm{y}= ( \bm{C_a} \circ \bm{F}(\mu) )^{-1} \ [\bm{m_B} \quad T_{\Sigma,d}]^T
\end{equation}
Where, $\bm{C_a}$ is the control allocation matrix described in Eq.\ref{mixer_matrix}, $\circ$ is the element-wise multiplication operator. The matrix $\bm{F}$ depends on the faulty actuator index. Considering no fault ($\mu =0 $), $\bm{F}$ is initialized as the matrix of ones ($\bm{F}=\bm{1}_{4\times4}$). If $\mu$ is non-zero, then depending on it's value the following elements of $\bm{F}$  are changed ($\pm$ denotes whichever exists).
 \begin{equation}
     \begin{split}
       F_{1\mu}=F_{2\mu}&=F_{4\mu}=0\\
      F_{3\mu}&= 1/k_{\tau}(\gamma_\mu)\\
      F_{3(\mu\pm2)}&=F_{4(\mu\pm2)}=0    
     \end{split}
 \end{equation}
Where $F_{ij}$ is the element in the $i^{th}$ row and $j^{th}$ column of $\bm{F}$. In case of no-fault, the Eq.\ref{CA} takes the form of standard control allocation. The resulting vector $\bm{y}$ will consist of the desired thrust that has to be generated by each actuator. During the faulty condition, $\bm{y}$ consists of the desired thrust for three working actuators and a desired torque for the actuator opposite to the failed one. The actuator commands that generate the desired thrust/torque are required.

Due to redundancy in the Heliquad's actuator, many actuator commands can generate the desired thrust. As discussed in Section.\ref{intro_section},  depending on the requirement, there are many ways to resolve this redundancy. However, in this paper, during fault-free upright and inverted flight, the propeller pitch angles are fixed, and the attitude is controlled purely by varying the motor RPM. The sign of $[T_i]_d$'s changes when mid-flight flipping is commanded. Hence, $\gamma_i$'s are also reversed. After the complete failure of an actuator, $\gamma$'s for the two opposite working actuators are fixed to a known value, and depending on their index, either roll or pitch is controlled by varying the RPM only. In contrast, both $\gamma$ and $\Omega$ are varied on the actuator located opposite to the failed one. This single actuator controls roll/pitch and yaw-rate simultaneously.
 
 Neural Networks (NN) are used to obtain the static map between the desired thrust/torque to actuator commands. Two different NN architectures are implemented in the Heliquad's control allocation. In the fault-free scenario, the first NN (NN1) approximates the relationship between $[T_i \quad \gamma_i]^T_d$ and $[ \Omega_i]_d$. There are four NN1's in total, one for every actuator. In faulty cases, the NN used for the actuator opposite to the failed one switches to another NN (NN2) that approximates the relationship between $[T_i \quad \tau_i]_d^T$ and $[\gamma_i \quad \Omega_i]_d^T$. The number of Neural Networks executed in control allocation at any given time is equal to the number of working actuators. Both the NN architectures have a single hidden layer, and the network parameters (weights and biases) are trained by the Bayesian regularization-backpropagation method on the experimental data collected on the load cell.  The number of neurons in the hidden layer is selected by the method described in \cite{NNidentification}. In practice, a saturation and rate limiter is cascaded with the output of NN-based control allocation to prevent the adverse effects of unbounded actuator commands, if any.

\section{Heliquad Prototype Description and Performance Evaluation} \label{heli_results_section}

In this section, parametric details of the VPP mechanism implemented on the Heliquad prototype are presented. The prototype's physical attributes and its performance across different flight modes are also examined in detail.

\subsection{ Variable-Pitch-Propeller Mechanism Prototype}

The complete VPP mechanism of the Heliquad is shown in Fig.\ref{mechanism_schematic_b}.  Commercially-Off-The-Shelf (COTS) BLDC motor is modified by replacing the existing shaft with a hollow one. The servo-motor is fixed beneath the BLDC motor. The links with revolute joints are connected to each other by a screw passing through the copper bush, providing a smooth interface. All the links, including the slot, are machined in Aluminium. The mechanism parameters are shown in Table.\ref{mech_params}

\begin{figure*}[htbp]
    \centering
    \begin{subfigure}[b]{0.3\textwidth}
    \includegraphics[width=\columnwidth]{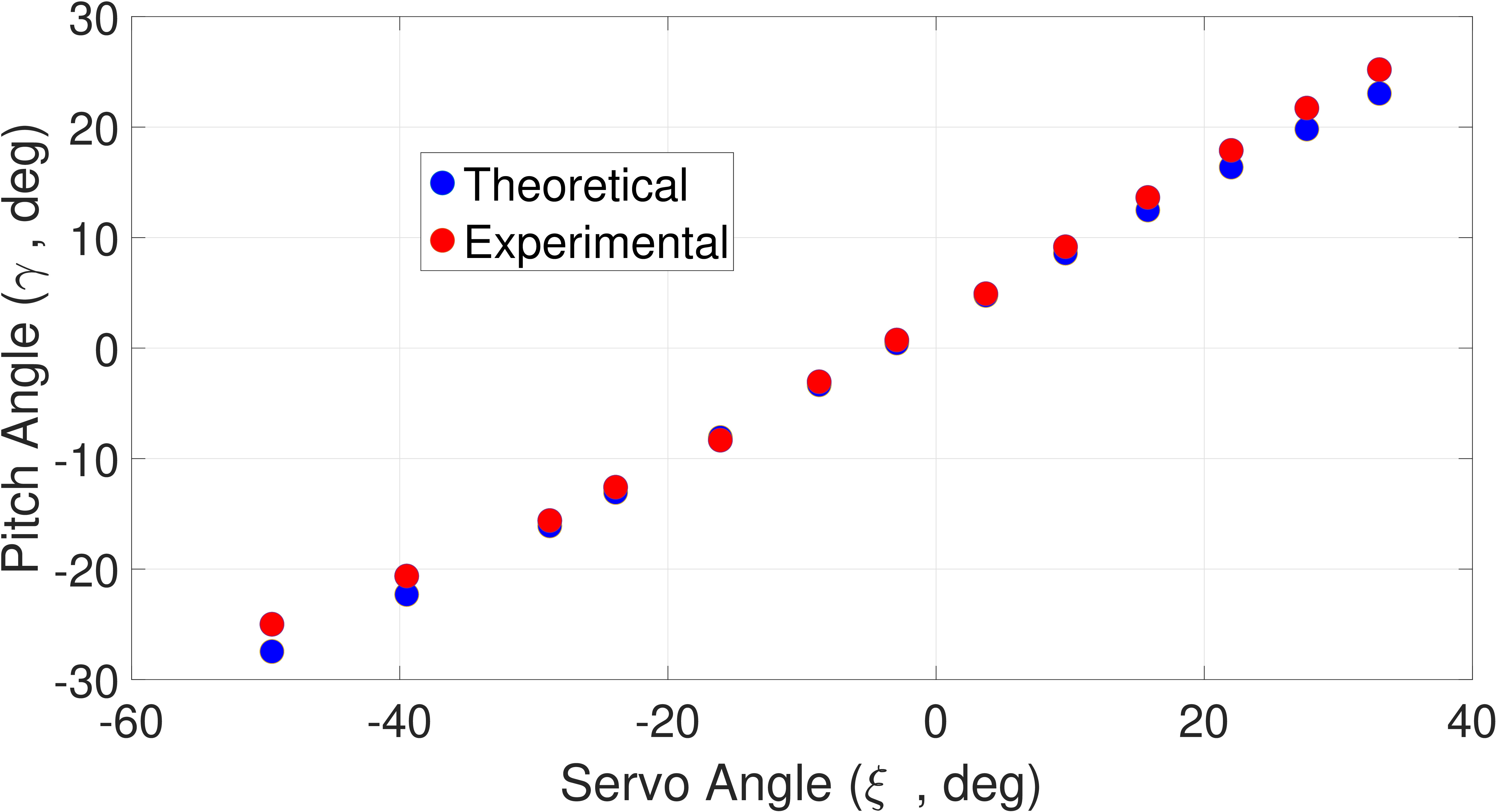}
    \caption{$\xi$ \  vs \ $\gamma$}
    \label{xi vs alpha}   
    \end{subfigure}
    \begin{subfigure}[b]{0.3\textwidth}
    \includegraphics[width=\columnwidth]{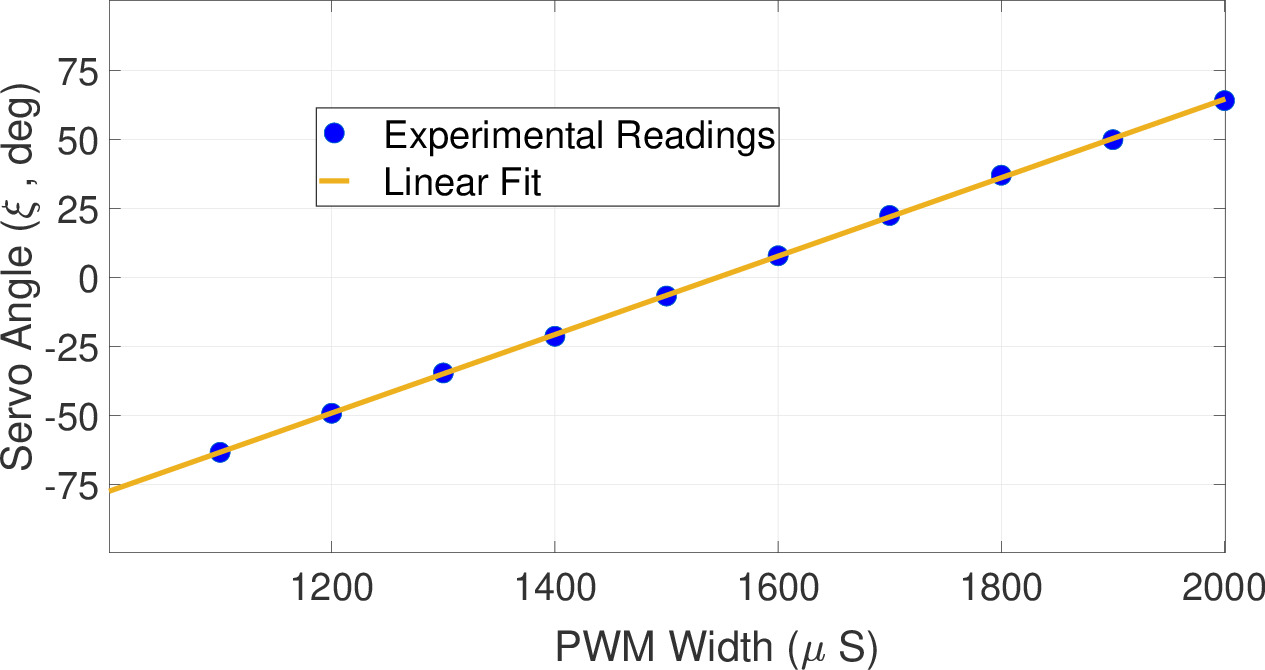}
    \caption{Servo PWM \  vs \ $\xi$}
    \label{servopwm vs xi}   
    \end{subfigure}
        \begin{subfigure}[b]{0.29\textwidth}
    \includegraphics[width=\columnwidth]{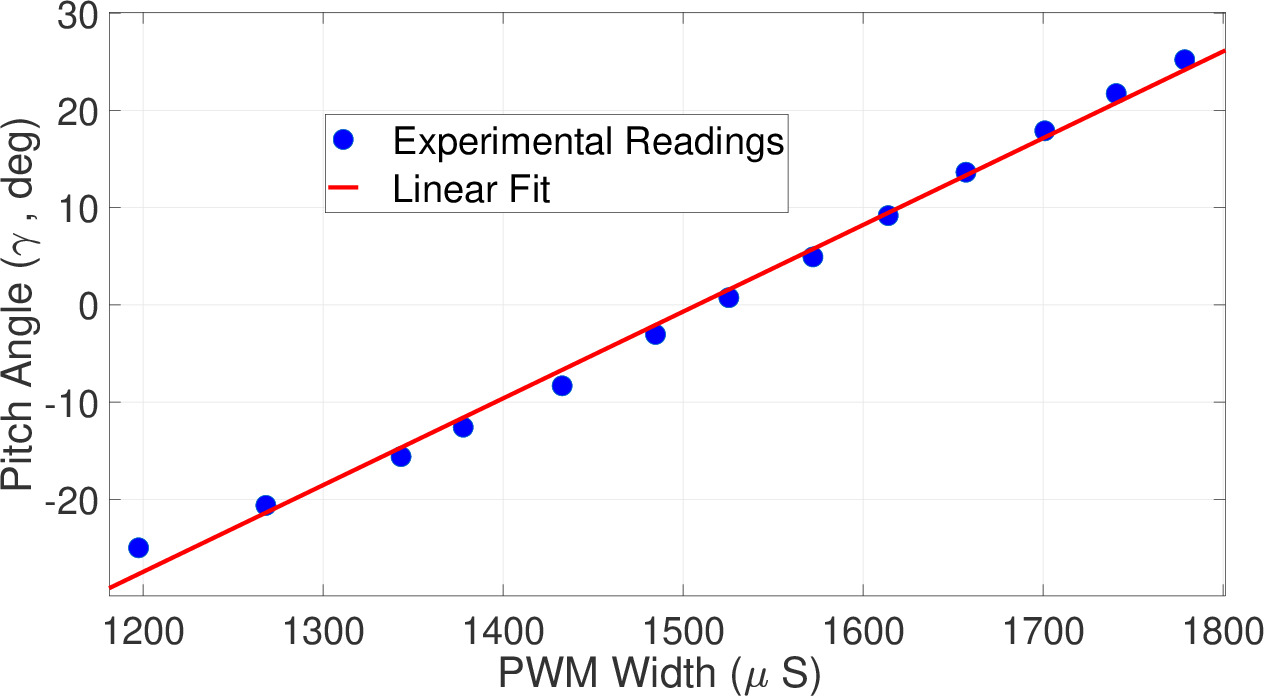}
    \caption{Servo PWM \  vs \ $\gamma$}
    \label{servopwm vs alpha}   
    \end{subfigure}
\caption{Input-Output relationship for the VPP mechanism. All the readings were taken on the same setup}
\end{figure*}

\begin{table}[htbp]
    \centering
    \caption{VPP mechanism parameters}
    \begin{tabular}{|m{0.2\columnwidth} | m{0.2\columnwidth}| m{0.2\columnwidth}| }
    \hline
    \textbf{Parameter} & \textbf{Value} & \textbf{Unit} \\
    \hline
    $n_B$ & 2 & -\\
    \hline
    $l_1$ & 55.29 & mm\\
    \hline
    $l_2$ & 7.98 & mm\\
    \hline
    $l_3$ & 6.73 & mm\\
    \hline
    $l_4$ & 8.58 & mm\\
    \hline
    $r$ & 5.6 & mm\\
    \hline
    $X_0$ & 0 & mm\\
    \hline
    $Y_0$ & 46.35 & mm\\
    \hline
    $\eta_1$ & 256.17 & deg\\
    \hline
   \end{tabular}
   \label{mech_params}
\end{table}

The values from Table.\ref{mech_params} are substituted in Eq.\ref{iorel} to estimate the input-output relationship. This theoretical data has to be validated against the experimentally measured values. A digital pitch gauge is used to measure the propeller pitch angle at different servo-motor angles. The motor was not spinning while measuring the pitch angle. As the propeller stalls at higher pitch angles \cite{cutler_2015}, the comparison for the values between $\pm30$ deg is shown in Fig.\ref{xi vs alpha}.

It can be seen that Eq.\ref{iorel} predicts the relationship with sufficient accuracy. The experimental readings are within 10\% of theoretical estimates for all the pitch angles except the reading near 0 degrees, where it is around 30\%. This may be due to the pitch gauge's limited precision, as the measured value is very small. In practice, the servo angle cannot be directly commanded. The servo-motor usually accepts the PWM signal ($\zeta$) with a width between 1000 and 2000 $\mu S$ corresponding to opposite extremities of the servo angle. The variation for the values in between is shown in Fig.\ref{servopwm vs xi}. The servo-angle readings were estimated from the images and validated by values from the internal potentiometer. The relationship is almost linear with a Root Mean Square Error (RMSE) of 0.13 deg. A linear fit may also be used to approximate the relation between the servo PWM and pitch angle, as shown in Fig.\ref{servopwm vs alpha}. Here, the RMSE is 0.3 deg. A similar input-output trend is demonstrated in \cite{hrishikeshavan}. Based on Fig.\ref{servopwm vs alpha}, the mathematical relation used in the Heliquad's controller is given by Eq.\ref{math_servopwm_vs_alpha}
\begin{equation}
\gamma_{deg} \ = \ 0.09\ ( \zeta \ - \ \zeta_0 )
\label{math_servopwm_vs_alpha}
\end{equation}
where, $\gamma_{deg}$ is the propeller pitch angle measured in degrees and $\zeta_0$ is the  PWM width for zero servo angle. Due to manufacturing imperfections, every servo-motor on the Heliquad may not be exactly the same. Hence, $\zeta_0$ may be different for every actuator. Similar to any BLDC motor, the RPM of the motors on the Heliquad is also controlled by a PWM signal.

\subsection{Heliquad Prototype}

\begin{figure}[h!]
 \begin{subfigure}{0.5\columnwidth}
  \includegraphics[width=\columnwidth]{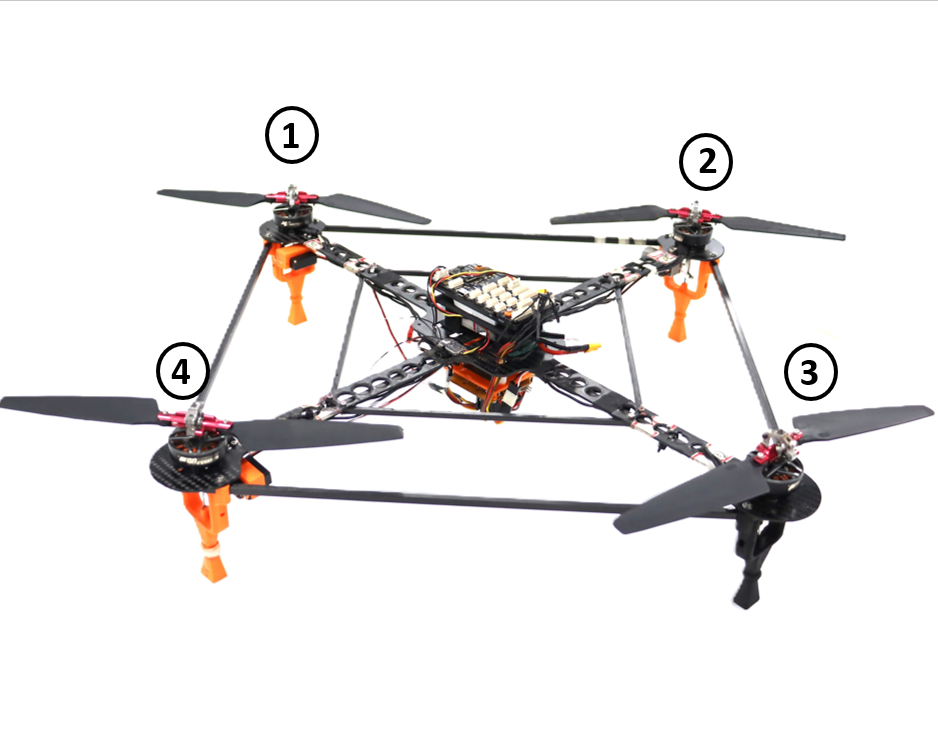} 
  \caption{Oblique view with actuator numbering. }
 \end{subfigure} 
 \hfill
 \begin{subfigure}{0.45\columnwidth}
  \includegraphics[width=\columnwidth]{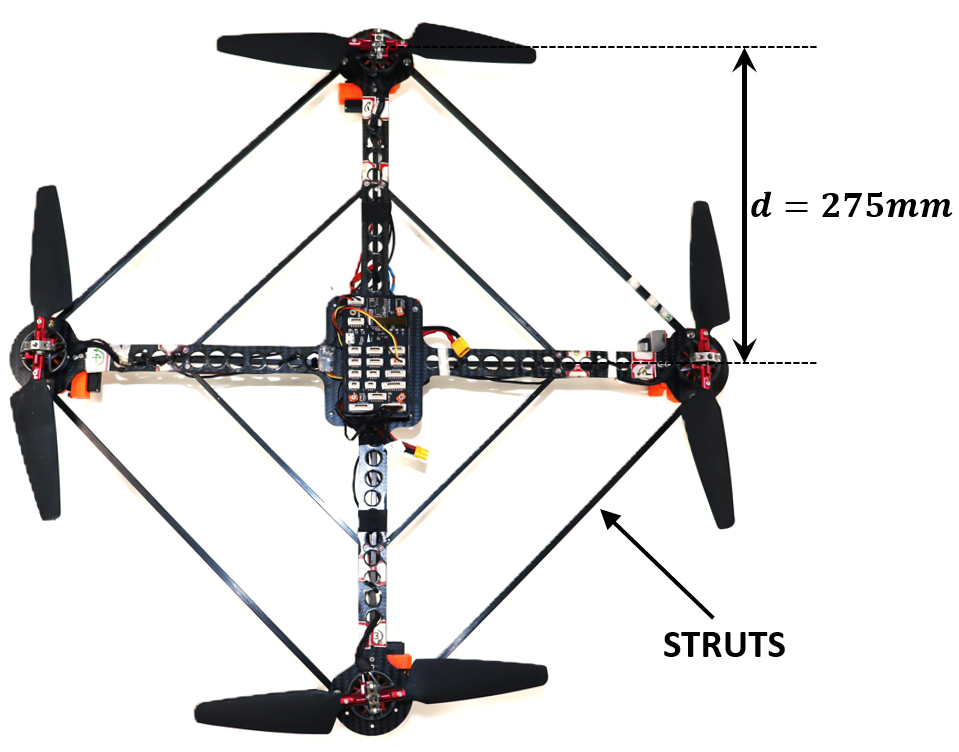}
  \caption{Top view. Struts play a crucial role in reducing vibrations. }
 \end{subfigure}
 \caption{ Heliquad prototype used for the experiments}
 \label{heli_proto}
\end{figure}

Apart from regular flights, the Heliquad prototype is expected to do mid-flight flipping and flight on three actuators. It is necessary for the prototype to be as lightweight as possible with sufficient stiffness to counter high vibrations, primarily induced by the rapidly spinning VPP mechanism and propellers. The airframe of the prototype is of '+' shape with the opposite motor centers 550 mm apart. The airframe is manufactured using carbon composites. The struts are connected to adjacent arms of the prototype, and they play a crucial role in preventing vibrations caused by frame-flexing \cite{frame-flex, cutlerthesis}. The Heliquad prototype built for the experiments is shown in Fig.\ref{heli_proto}.

The rotors are assembled by changing the shaft of the COTS 1300KV BLDC motor. The autopilot onboard has a hardware architecture similar to open-source 'Pixhawk FMUv5' \footnote{\url{https://github.com/pixhawk/Hardware}}. The attitude controller was executed at 1000Hz for all the experiments. The attitude reference commands were sent over a hand-held radio transmitter. A 4-cell Lithium-Polymer (Li-Po) battery powers all the actuators and electronics onboard. The detailed weight distribution of the prototype is given in Table.\ref{proto_weight}. Note that VPP-specific hardware amounts to only 15\% of overall weight.

\begin{table}[h]
    \centering
    \begin{tabular}{|c|c|c|}
    \hline
        \textbf{SR.No} & \textbf{Component}  & \textbf{Weight} (gms)\\
        \hline
         1 &  BLDC motor (x4)& 192\\
         \hline
         2 &  VPP mechanism (x4)& 44 \\
         \hline
         3 &  servo-motor (x4)& 48\\
         \hline
         4 &  Propeller blades (x8)& 48\\
         \hline
         5 &  Airframe& 134\\
         \hline
         6 & Battery & 75\\
         \hline
         7 &  Electronics and wires & 85\\
         \hline
           & \textbf{Total} & 626\\
           \hline
    \end{tabular}
    \caption{Weight Distribution of the prototype}
    \label{proto_weight}
\end{table}

\begin{figure}[h!]
 \begin{subfigure}{0.48\columnwidth}
  \includegraphics[width=\columnwidth]{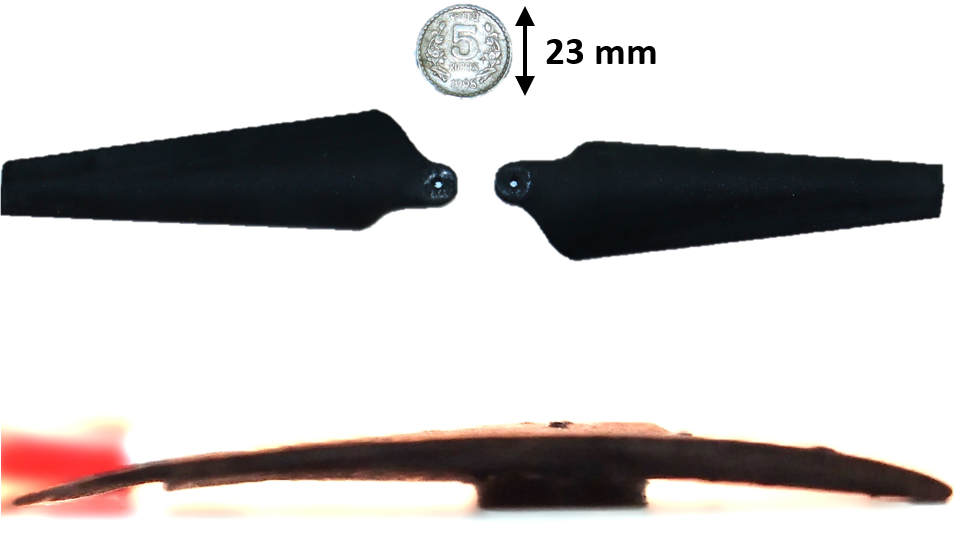} 
  \caption{Propeller planform (top), and cambered airfoil (bottom) }
 \end{subfigure} 
 \hfill
 \begin{subfigure}{0.48\columnwidth}
  \includegraphics[width=\columnwidth]{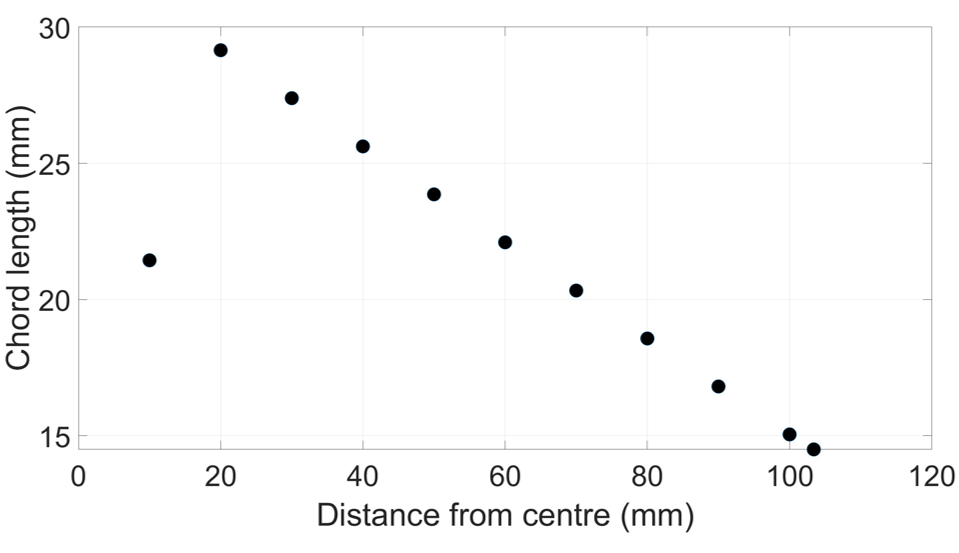}
  \caption{Chord length variation \\ }
 \end{subfigure}
 \caption{Heliquad propeller. Cambered airfoil plays a crucial role in achieving full-attitude control on three actuators.}
 \label{prop_proto}
\end{figure}

 The propeller in the Heliquad plays a crucial role in Heliquad's ability to fly with full-attitude control on three actuators. It is shown in \cite{kulkarni} that only cambered airfoil propeller blades can feasibly generate enough torque at zero thrust to enable yaw-rate control. Towards it, the prototype propeller blade has a planform similar to the COTS  propeller. However, the airfoil is modified to a cambered Eppler-63 \footnote{\url{http://airfoiltools.com/airfoil/details?airfoil=e423-il}}.  The propeller and its chord length variation are shown in Fig.\ref{prop_proto}. The blades are untwisted
 and 3D-printed using a Fused-Deposit-Modelling (FDM) method. Cambered airfoil propellers produce higher pitching moments compared to their symmetric airfoil counterparts. Using the chord length variation in BEMT equations, the maximum estimated moment of 0.0105 N-m generated by the prototype propeller is obtained at $\gamma$ = 19 deg and $\Omega$ = 15000 RPM. Substituting this value and link parameters from Table.\ref{mech_params} in Eq.\ref{prop_moment}  leads to the estimated torque of 0.06 Kg-cm acting on the servo-motor. The servo-motors on the Heliquad have a maximum torque of 4.8 Kg-cm at 6V.

A single actuator was tested on the load cell.  Various performance parameters, including thrust and torque, were recorded for the range of propeller pitch angles and motor RPMs. The pitch angle was varied from -25 to +25 deg, and for every pitch angle, the RPM was increased from minimum to maximum in multiple random steps. A total of 320 data points were collected to train the NNs employed in the Control Allocation. Both NN architectures have 15 neurons with the tanhyperbolic activation function in the hidden layer. The dataset was randomly divided into a 70:30 ratio for training and testing, respectively. The data was normalized between [0,1]. After training for 1000 epochs using the Bayesian-regularization method, the test data MSE for NN1 was 1.3\%, and for NN2, it was 1.8\%.

 \subsection{Experimental flight tests}

Experimental flight tests were conducted on the prototype to demonstrate mid-flight flipping, full-attitude controlled flight on three actuators, and to validate the tracking performance of the attitude controller. The controller parameters are given in Table.\ref{controller_params}. While testing, the prototype was free and not supported to the ground in any means. The related videos are included in the supplementary material.

\begin{table}[htbp]
    \centering
    \begin{tabular}{|c|c|c|}
    \hline
        \textbf{SR.No} & \textbf{Parameter}  & \textbf{Value} \\
        \hline
         1 &  $\bm{K_a}$ & diag\ [8 \ 8 \ 0]\\
         \hline
         2 & $\bm{K_p}$ & diag [0.25 \ 0.23 \ 0]\\
         \hline
         3 & $\bm{K_I}$ & diag\ [0.35 \ 0.35 \ 0.2] \\
         \hline
         4 & $\bm{K_D}$ & diag\ [0.0003 \ 0.0003 \ 0] \\
         \hline
    \end{tabular}
    \caption{Controller Parameters. The same parameters are used for all experiments.}
    \label{controller_params}
\end{table}

\subsubsection{Mid-flight Flipping} \label{sec_Flipping}

In this experiment, the mission started with the Heliquad taking off from rest on the ground in an upright state. After it reached a certain altitude, $\sigma (t)$ was toggled by a switch on the hand-held radio transmitter to command inverted flight. After flipping, the inverted state was maintained, and basic maneuvers to validate attitude reference tracking were conducted before toggling $\sigma(t)$ again to command upright flight. Multiple flips were conducted before landing in an upright condition. 

\begin{figure}[htbp]
    \centering
    \includegraphics[width=0.95\columnwidth]{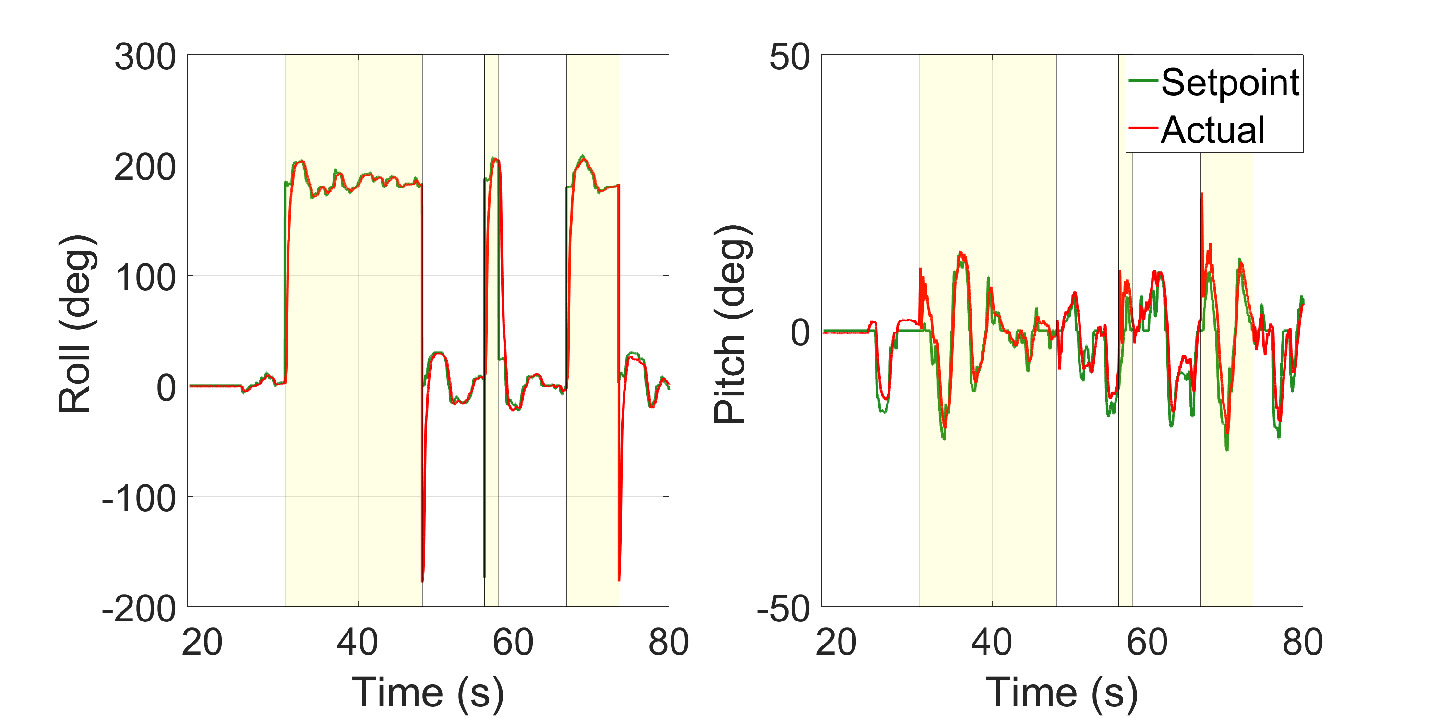}
    \caption{Pitch and Roll tracking performance. Yellow background denotes inverted flight}
    \label{tvsattitude_flip}
\end{figure}

\begin{figure}[htbp]
    \centering
    \includegraphics[width=\columnwidth]{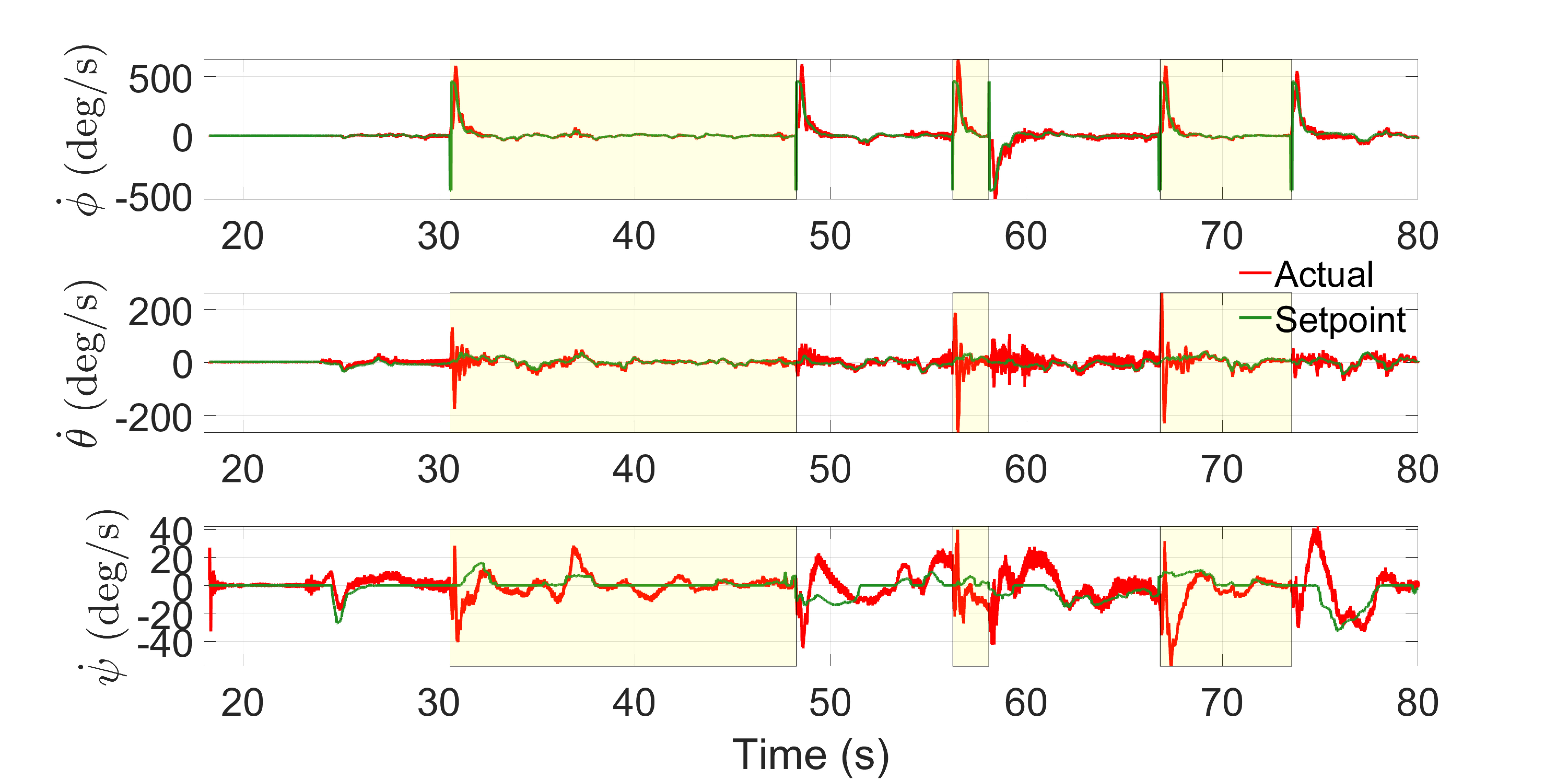}
    \caption{Attitude rate tracking performance. The yaw-rate is directly commanded.}
    \label{tvsrates_flip}
\end{figure}

The attitude tracking results are shown in Fig.\ref{tvsattitude_flip}. For better representation, the values for roll angle in the inverted flight are transformed to [0 360] degrees. It can be seen that the controller tracks the references well for both upright and inverted conditions. The RMSE for the roll is 0.72 deg, and for pitch, it is 0.93 deg. The rate controller performance is shown in Fig.\ref{tvsrates_flip}. The results show satisfactory tracking performance. At very high roll rates that occur during flipping, there is a simultaneous divergence in pitch and yaw-rates. It may be due to adverse inertial-roll coupling phenomena. However, the controller is able to quickly control the rates without causing instability.

\subsubsection{Flight on three actuators} \label{sec_3mot}

\begin{figure}[htbp]
    \centering
    \includegraphics[width=\columnwidth]{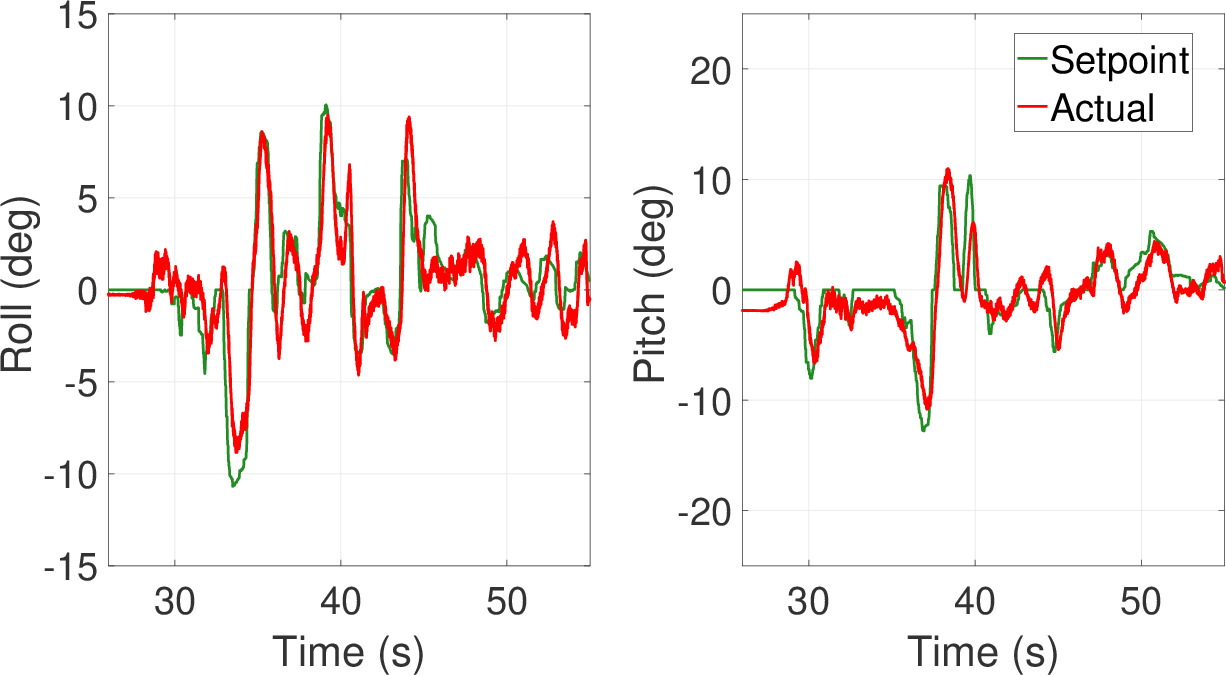}
    \caption{Pitch and Roll tracking on three actuators}
    \label{tvsattitude_3mot}
\end{figure}

\begin{figure}[htbp]
    \centering
    \includegraphics[width=0.95\columnwidth]{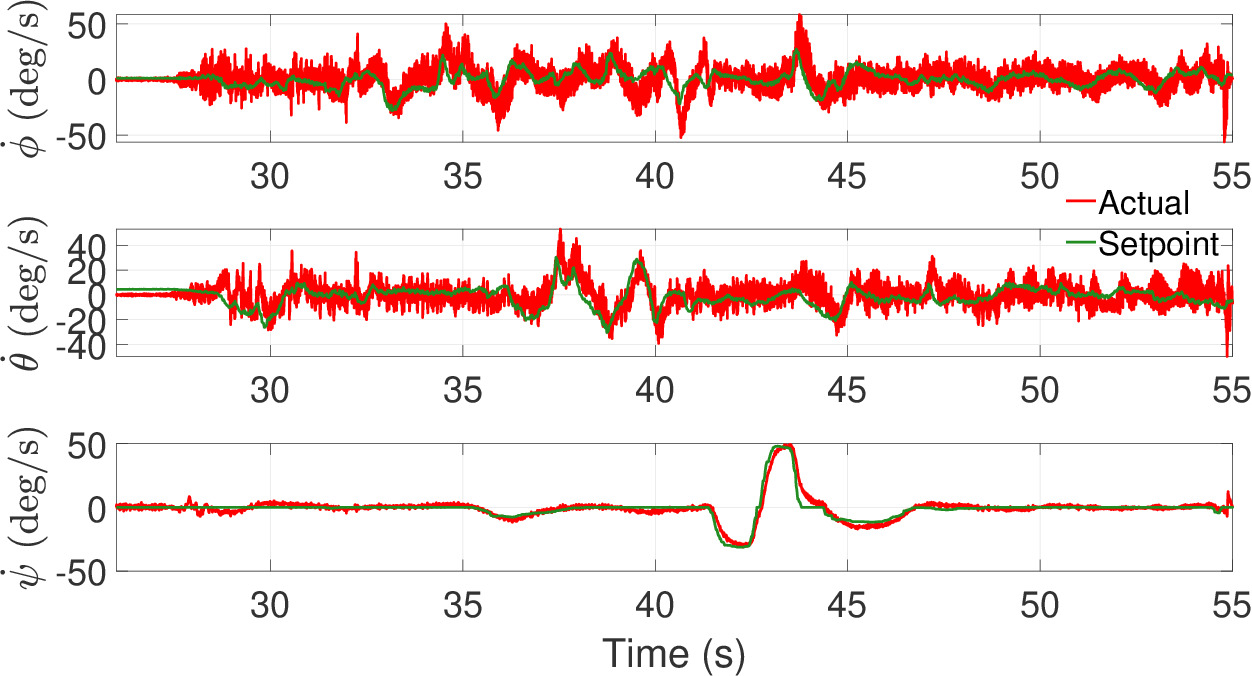}
    \caption{Rate tracking on three actuators. yaw-rate is commanded externally.}
    \label{tvsrates_3mot}
\end{figure}

In this experiment, the full-attitude control on only three working actuators is demonstrated. Without any loss in generality, motor number 4 was turned off before taking off in an upright position. The controller is initialized with $\mu=4$ in Eq.\ref{CA}. Fig.\ref{tvsattitude_3mot} shows the attitude tracking performance. The RMSE in roll is 0.98 deg, and in pitch it is 1.12 deg. The performance has degraded slightly compared to nominal experiments.  It is suspected that higher vibrations may be causing the error in the attitude estimation. 
Fig.\ref{tvsrates_3mot} shows the rate tracking performance. All the attitude rates are tracked well. It can be seen that, unlike fixed-pitch quadcopters, Heliquad is able to control the yaw-rate only on three of its working actuators. It's also worth noting that a single actuator, in this case actuator 2, is able to track the roll-rate and yaw-rate setpoints simultaneously. This ability of the Heliquad will be crucial to ensure a safe recovery and mission continuation during the mid-flight actuator failure crisis.

\begin{figure}[htbp]
    \centering
    \includegraphics[width=0.95\columnwidth]{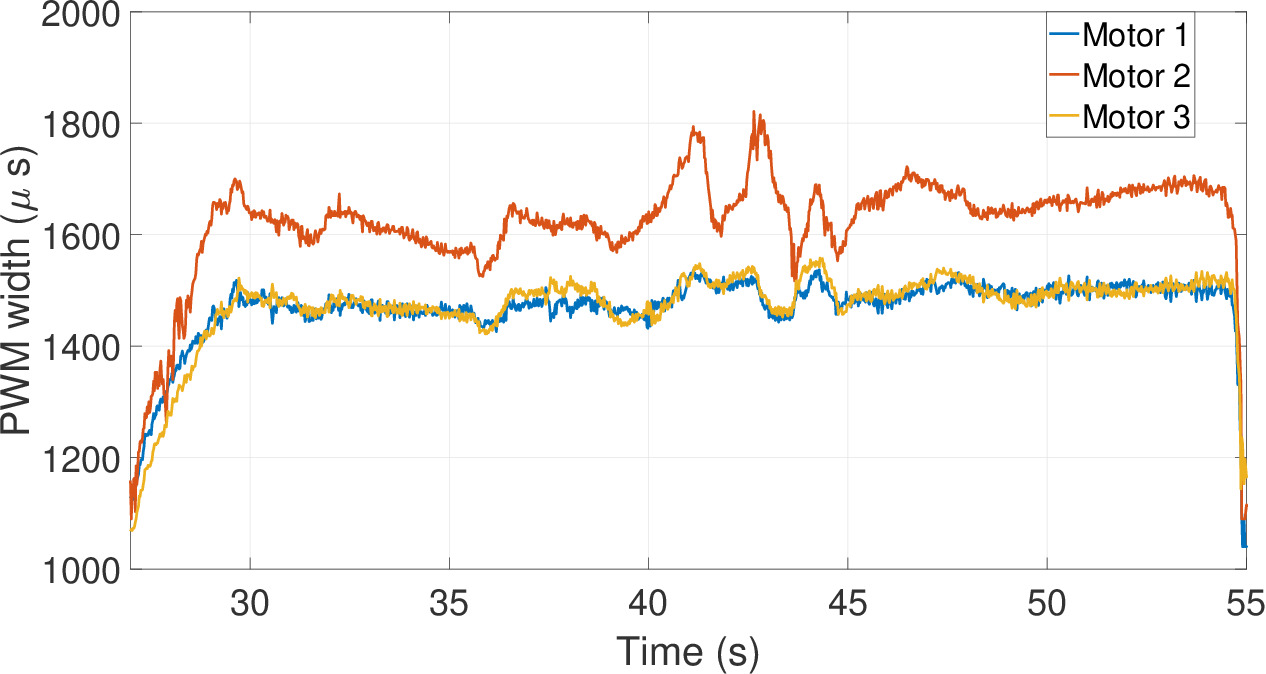}
    \caption{Commanded pulse width variation for Heliquads BLDC motors}
    \label{tvsesc_3mot}
\end{figure}

\begin{figure}[htbp]
    \centering
    \includegraphics[width=0.95\columnwidth]{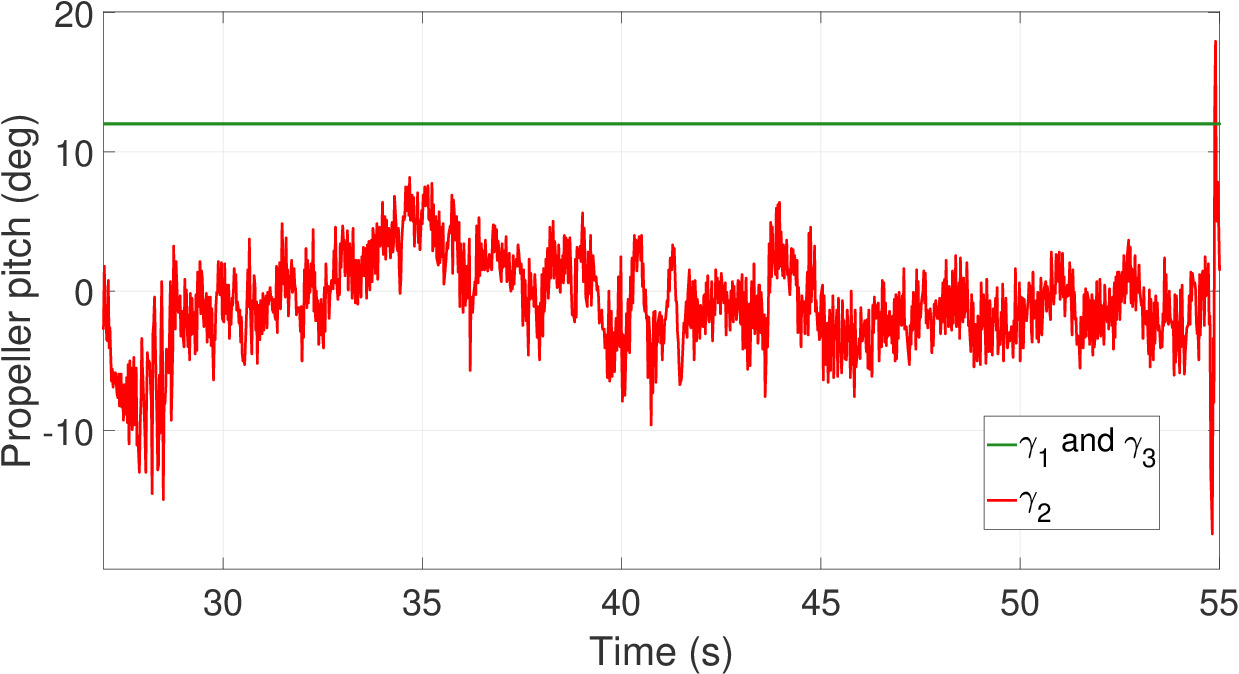}
    \caption{Commanded propeller pitch angle variation for Heliquads VPP mechanism.}
    \label{tvspitch_3mot}
\end{figure}

To get better insight into the working of the controller,  motor PWM commands are shown in Fig.\ref{tvsesc_3mot}. The desired RPM computed by NN-based control allocation is bounded and within the limits throughout the experiment. As expected, to counter the cumulative torque generated by the thrust-producing rotors, $\Omega_2$ is consistently higher than the RPMs of two opposite working actuators. Almost overlapping $\Omega_1$ and $\Omega_3$ over a wide range indicate nearly the same $\gamma_1$ and $\gamma_3$ in flight. As there is no provision to measure the propeller pitch angle while flying, Eq.\ref{math_servopwm_vs_alpha} is used to estimate the commanded pitch angle. The variation is shown in Fig.\ref{tvspitch_3mot}. As per the strategy, $\gamma_1$ and $\gamma_3$ are fixed to 12 deg each. $\gamma_2$, which is also the output of NN2, is bounded and always near zero-thrust pitch angle (-2 deg).

\subsubsection{Midflight Motor Failure}

\begin{figure}[ht]
    \centering
    \includegraphics[width=0.95\columnwidth]{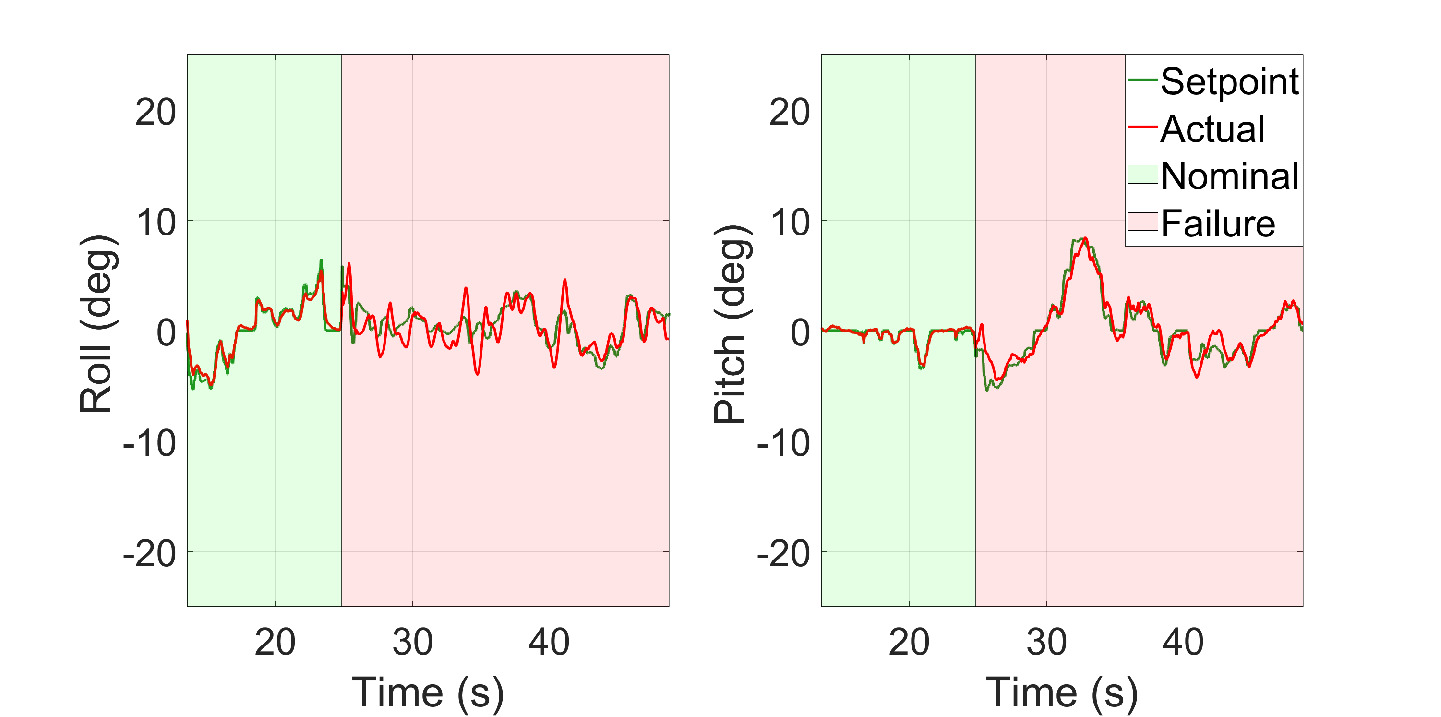}
    \caption{Roll and Pitch variation before and after the actuator failure.}
    \label{tvsattitude_failure}
\end{figure}

\begin{figure}[htbp]
 \begin{subfigure}{0.3\columnwidth}
  \includegraphics[width=\columnwidth]{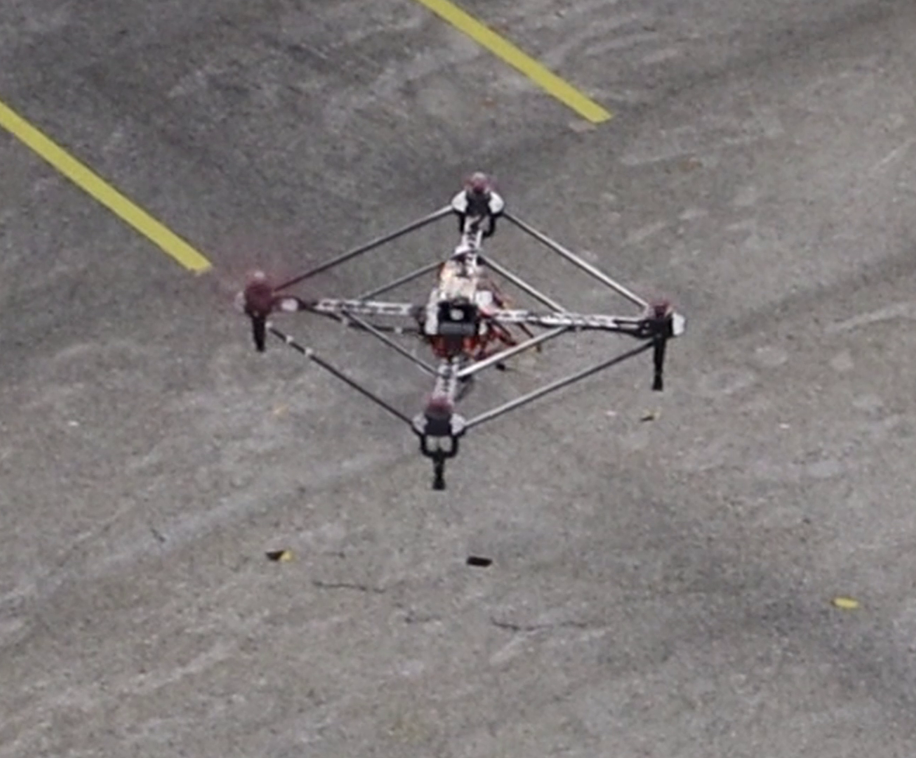} 
  \caption{Nominal flight. }
 \end{subfigure} 
 \hfill
 \begin{subfigure}{0.3\columnwidth}
  \includegraphics[width=\columnwidth]{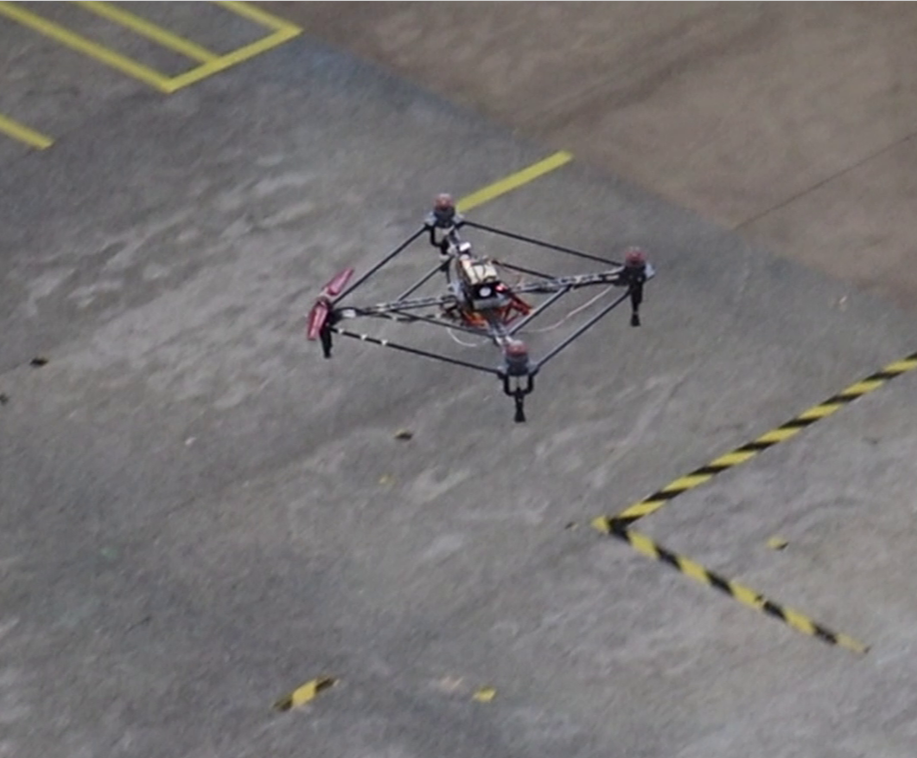}
  \caption{Motor failure. }
 \end{subfigure}
 \hfill
 \begin{subfigure}{0.31\columnwidth}
  \includegraphics[width=\columnwidth]{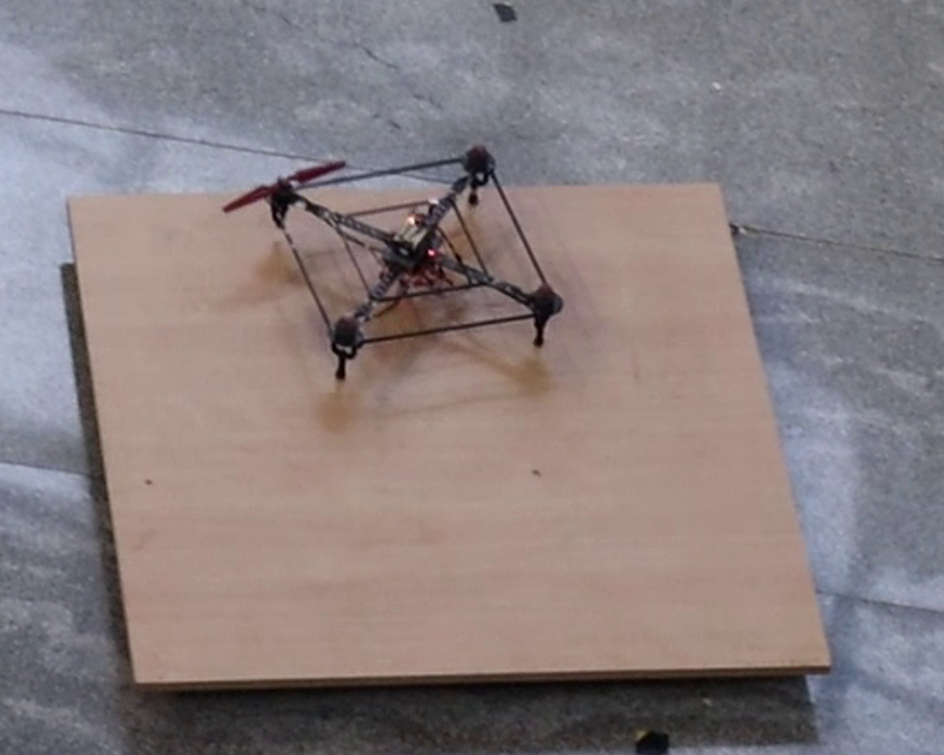} 
  \caption{Safe Landing.}
 \end{subfigure} 
 \caption{ Heliquad prototype lands precisely on the target with full-attitude control despite the complete failure of an actuator. Note, Motor with the red color propeller is switched off.}
 \label{heli_safe_land}
\end{figure}

\begin{figure}[ht]
    \centering
    \includegraphics[width=0.95\columnwidth]{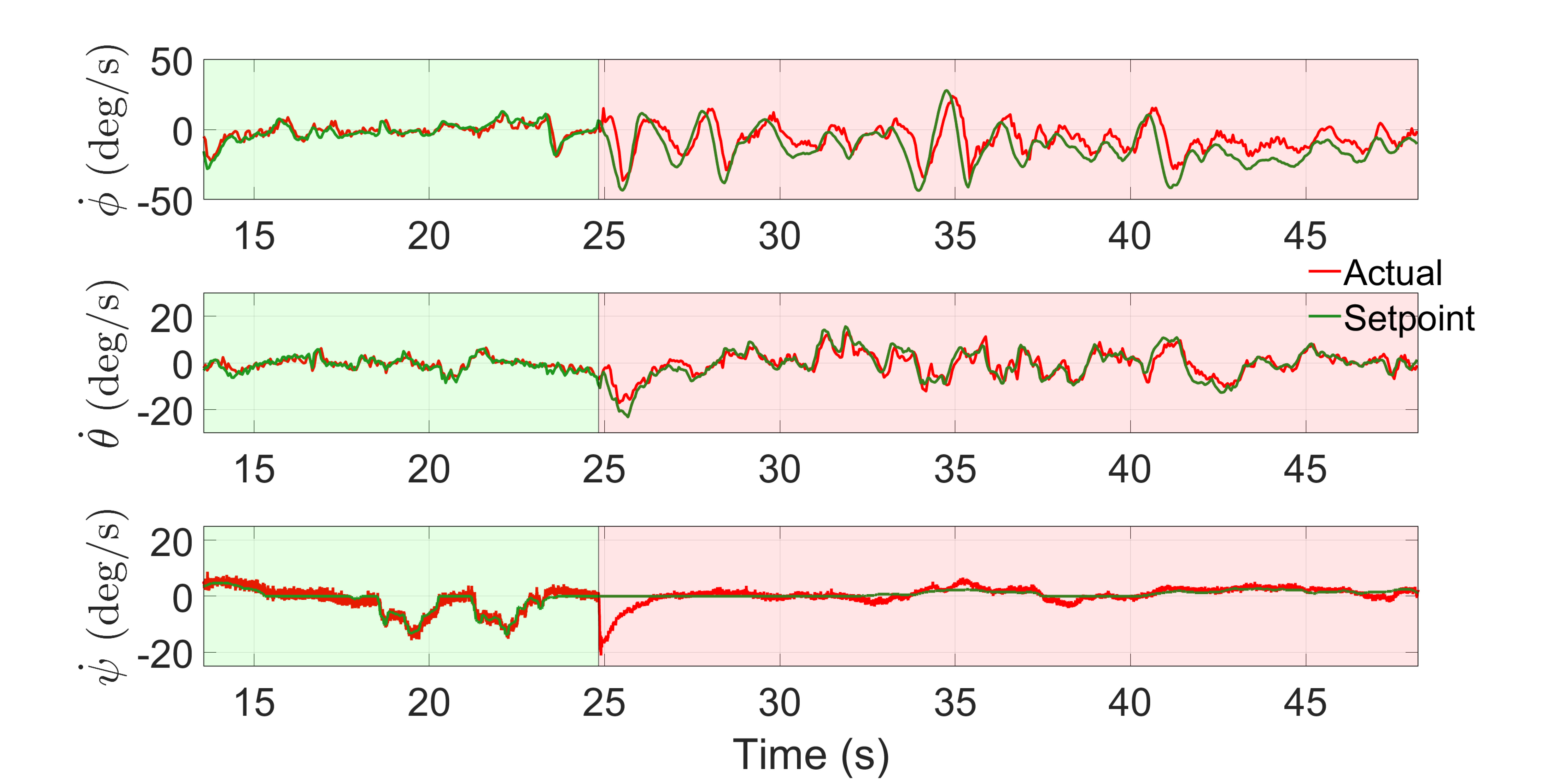}
    \caption{Rate tracking variation before and after the actuator failure.}
    \label{tvsrates_failure}
\end{figure}

In this experiment, the Heliquad takes off from the ground with all (four) working actuators. However,  motor 4 is switched off after reaching a certain altitude to replicate the complete failure of an actuator. Here, the control allocation is initialized with $\mu =0$, and it reconfigures by setting  $\mu=4$ after failure. The FDI delay time is not considered. The controller's tracking performance before and after the failure is shown in Fig.\ref{tvsattitude_failure} - \ref{tvsrates_failure}. The region with a green backdrop indicates the nominal flight with all working actuators, and the red backdrop represents the failed actuator scenario. The motor is turned off approximately at $t=25 s$.

It can be seen that the tracking performance in Fig.\ref{tvsattitude_failure} is in alignment with the results obtained in section \ref{sec_Flipping} and \ref{sec_3mot}. The tracking performance after failure, especially for roll angle, deteriorates compared to the nominal case. The degraded attitude estimates due to increased vibrations and higher uncertainty in actuator commands for roll may be the reason for degraded performance. However, even after the failure, the Heliquad recovered safely, and it was responsive enough (to attitude commands) to land at the desired location, as shown in Fig.\ref{heli_safe_land}.

\begin{figure}[ht]
    \centering
    \includegraphics[width=0.95\columnwidth]{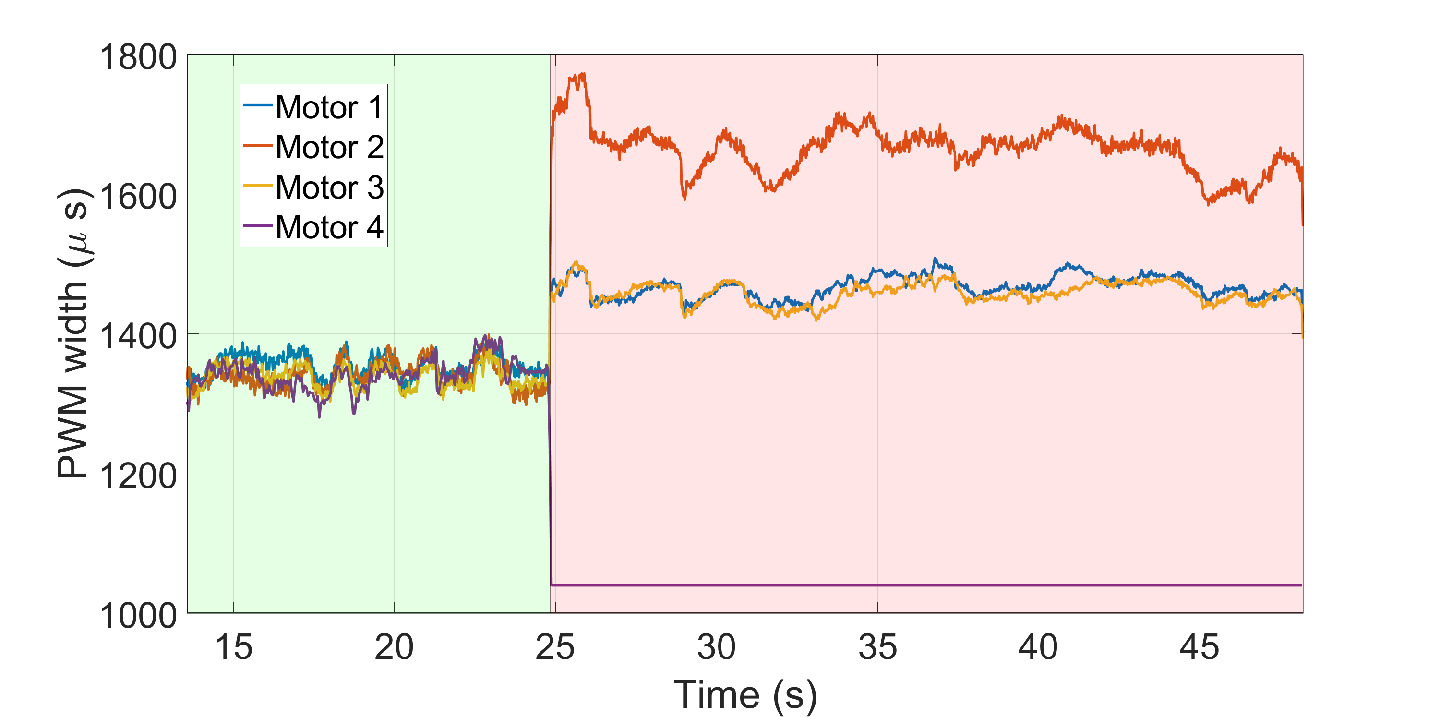}
    \caption{Motor PWM commands variation before and after the actuator failure.}
    \label{tvsesc_failure}
\end{figure}

The rate tracking performance is shown in Fig.\ref{tvsrates_failure}. As expected, the tracking performance of roll rate and pitch rate after failure has slightly worsened. However, after failure, there is a sudden divergence in yaw-rate, which could be due to torque imbalance resulting from the time delay of the BLDC motor to reach a higher RPM. Nonetheless, the controller quickly stabilizes and tracks the yaw-rate on three working actuators. The variation of BLDC motor PWM commands throughout the mission is shown in Fig.\ref{tvsesc_failure}

\begin{figure}[ht]
    \centering
    \includegraphics[width=0.95\columnwidth]{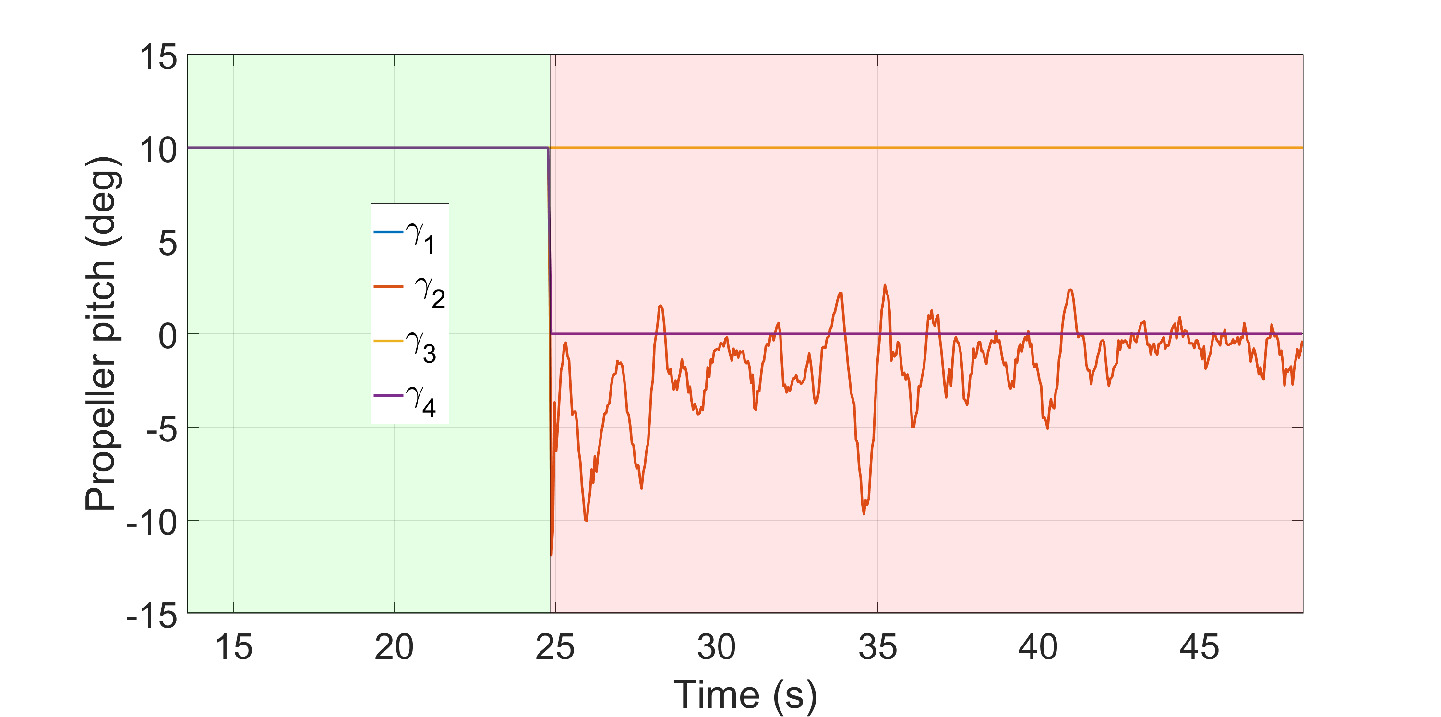}
    \caption{Propeller pitch angle commands variation before and after the actuator failure.}
    \label{tvspitch_failure}
\end{figure}

Before the failure, as the propeller pitch angle is commanded the same for all actuators, the BLDC motor RPM variation almost overlaps. After failure and control allocation reconfiguration, motors 1 and 3 commanded RPM increase to balance the weight of the Heliquad. Motor 2 commanded RPM increases significantly to balance the combined torque of opposite working actuators. The commanded propeller pitch variation is shown in Fig.\ref{tvspitch_failure}. After failure, $\gamma_1$ and $\gamma_3$ continue to maintain the same value. However, $\gamma_2$ varies near zero-thrust value to track the roll and yaw-rate simultaneously.
The absence of a high yaw-rate after failure ensures that the Heliquad can continue the mission and be landed safely, even in a manually piloted flight. In reduced-attitude control, due to very high yaw-rates, it may be difficult for a human pilot to maneuver the quadcopter precisely in a cluttered environment without position feedback, which in practice is commonly encountered in  GPS-denied environments.

Although the cambered airfoil propellers are efficient in producing positive thrust and generate significantly more torque at zero-thrust pitch angles, they are not efficient in producing negative thrusts. According to the load-cell-based propeller experimental data, for the flight on three actuators in an upright condition, the two opposite lifting propellers generate a hover thrust of 315 grams each by producing 0.052 N-m of torque. The cumulative 0.104 N-m is balanced out by motor 2 spinning at around 60\% throttle (approximately 7500 RPM). However, to generate -315 grams of thrust, the propeller produces 0.23 N-m torque. The combined 0.46 N-m is beyond the maximum zero-thrust torque of 0.195 N-m. Hence, the full-attitude equilibrium under complete failure of one motor is infeasible on the Heliquad's prototype in inverted flight. However, usual reduced-attitude techniques may be implemented to recover from mid-flight failure or full-attitude control may be regained by flipping to an upright state.

\section{Conclusion} \label{conclusion_section}

This paper presented the design and analysis of the Variable-Pitch-Propeller (VPP) mechanism and its implementation on a quadcopter called Heliquad to demonstrate unique flying characteristics. The input-output relation for the VPP mechanism was derived and experimentally found to be well approximated by a linear fit with RMSE of 0.3 deg. Various singularities of the mechanism were analyzed. Selecting appropriate link lengths ensured that these singularities did not fall within the operating pitch angle workspace. For proper actuator sizing, the expression for the minimum servo-motor torque required to enable pitch change was derived. The Heliquad is controlled by a unified cascaded PID attitude and rate controller followed by a unique Neural-Network-based control allocation. A prototype was built, and experimental free-flight tests were conducted to validate the utility of the mechanism and the controller's tracking performance. First, a mid-flight flipping experiment was conducted. The tracking performance of the controller was found to be excellent, with the RMSE values in roll and pitch at 0.72 and 0.93 degrees, respectively. The Heliquad prototype was then flown with one of its motors switched off. It was found that unlike standard quadcopters, where the yaw-rate control has to be sacrificed, the controller on Heliquad tracks the full-attitude references with only three working actuators in the upright state. Finally, the mid-flight actuator failure experiment demonstrated the Heliquad's ability to maneuver safely and land precisely. This novel capability was achieved by employing a cambered airfoil in the propeller blades. However, due to the cambered airfoil's inefficiency in producing negative thrust, it was found that the hover-equilibrium does not exist for the failure case in inverted flight. The outputs of the Neural-Network-based controller were bounded for all the experiments.

In future works, the effects of FDI delay time on the safe recovery of Heliquad in mid-flight actuator fault scenarios will be experimentally investigated.

\bibliographystyle{ieeetr}
\bibliography{Bibliography}

\end{document}


\maketitle

\section{Detailed derivation of mechanism input-output relationship}

Continuing from Fig.2 and Eq.7 in the main text one can write
\begin{equation}\label{PQeta}
 P \ cos\ \gamma \ + \ Q \ sin\ \gamma \ = \ A   
\end{equation}
Dividing Eq.\ref{PQeta} by $\sqrt{P^2 \ + \ Q^2}$

\begin{equation}\label{inter}
    \frac{P}{\sqrt{P^2 \ + \ Q^2}} cos\gamma \ + \ \frac{Q}{\sqrt{P^2 \ + \ Q^2}}sin\gamma \ = \ \frac{A}{\sqrt{P^2 \ + \ Q^2}}
\end{equation}

Eq.\ref{inter} can be written as
\begin{equation} \label{ccss}
  cos\nu cos\gamma \ + \ sin\nu sin\gamma \ =   \frac{A}{\sqrt{P^2 \ + \ Q^2}}
\end{equation}

where $\nu \ = \ \tan^{-1} \left(\frac{Q}{P}\right)$. Simplifying Eq.\ref{ccss},  we can get two solutions

\begin{equation}
    cos ( \nu \ - \ \gamma) \ = \ \frac{A}{\sqrt{P^2 \ + \ Q^2}}
\end{equation}
or
\begin{equation}
    cos (  \gamma \ - \ \nu) \ = \ \frac{A}{\sqrt{P^2 \ + \ Q^2}}
\end{equation}
Further simplification leads to
\begin{equation}
\begin{split}
       \gamma \ &= \  \nu \ \pm \ cos^{-1} \left(\frac{A}{\sqrt{P^2 \ + \ Q^2}}\right) \\
 &=  \  \tan^{-1} \left(\frac{Q}{P}\right) \ \pm \ cos^{-1} \left(\frac{A}{\sqrt{P^2 \ + \ Q^2}}\right)
\end{split}
\end{equation}
Therefore, the mechanism has two branches. The branch described by Eq.9 in the main text is assembled for the Heliquad prototype.

\section{Propeller pitching moment derivation}

\begin{figure}[t]
    \centering
    \includegraphics[width=\columnwidth]{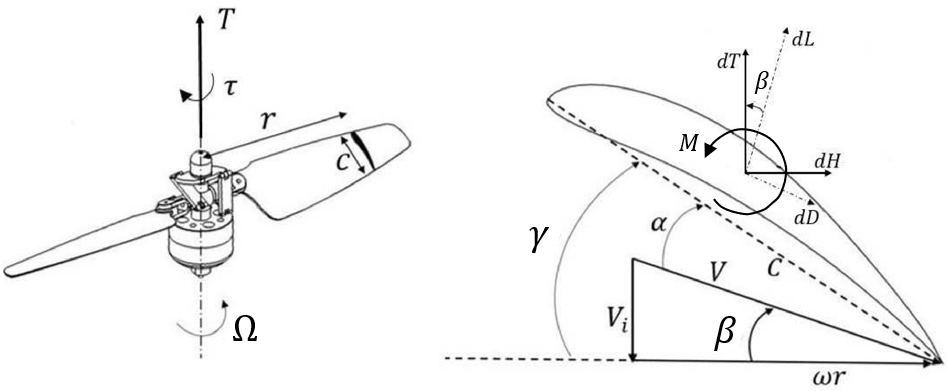}
    \caption{Propeller parameters (right). Propeller crosssection parameters (left). $r$ is the distance of the element from the center of rotation, $\alpha$ is the  angle of attack. }
    \label{bemt}
\end{figure}

As per Fig.\ref{bemt}, the elemental thrust generated by a propeller section can be estimated by Blade Element Theory (BET) as
 \begin{equation}
    \begin{split}
  dT &=  dL cos(\beta) - dD sin(\beta)   \\
  & = \frac{n_B}{2}\rho V^2 (C_l(\alpha) cos\beta- C_d(\alpha) sin\beta) c(r) dr
  \label{dtdl}
    \end{split}
 \end{equation}

where

 \begin{equation}\label{alphaphi}
   \begin{split}
        V^2 &= V_i^2+ (\Omega r)^2 \\
        \alpha &= \gamma-\beta = \gamma-tan^{-1}\frac{V_i}{\Omega r}    
   \end{split}  
\end{equation}

In Eq.\ref{dtdl}, $\rho$ is the density of surrounding air (Assumed $1.22$ Kg/cubic-m). $C_l$ and $C_d$ are the lift and drag coefficients of the airfoil. $c(r)$ is the chord length variation and depends on the propeller design. $V_i$ is the unknown induced velocity. To estimate $V_i$, the momentum theory is commonly used. Based on the conservation of linear momentum, it gives,

\begin{equation}
dT=4\pi \rho  V_i^2 r dr\\
\label{dtmomentum}
\end{equation}

Eq.\ref{dtdl} and Eq.\ref{dtmomentum} together is called the Blade Element Momentum Theory (BEMT). $V_i$ can be iteratively calculated for given $\gamma$ and $\Omega$ by equating Eq.\ref{dtdl} and Eq.\ref{dtmomentum}. Substituting $V_i$ in Eq.\ref{dtdl} and integrating it throughout the blade span gives the thrust generated by the propeller.
Similarly, the pitching moment generated by $n_B$ blades can be written as,

\begin{equation}
    M_{prop} \ = \  \int_{r_{min}}^{r_{max}} \frac{n_B}{2} \rho V^2 \ C_m(\alpha) \ c(r)^2 \ dr
\end{equation}